\documentclass[12pt,preprint]{aastex}

\def\deg{^\circ}

\def\be{\begin{equation}}
\def\ee{\end{equation}}

\def\cm{\hbox{\,cm}}
\def\yr{\hbox{\,y}}

\def\bfj{{\bf j}}
\def\bfe{{\bf e}}
\def\bfr{{\bf r}}
\def\bfR{{\bf R}}
\def\bfT{{\bf T}}

\def\half{{\textstyle{1\over2}}}

\def\bfL{{\bf L}}
\def\bfn{{\bf n}}
\def\bfr{{\bf r}}

\def\bfR{{\bf R}}

\def\bfv{{\bf v}}
\def\bfu{{\bf u}}

\def\bft{{\bf t}}

\def\bnabla{\mbox{\boldmath $\nabla$}}

\def\bfJ{{\bf J}}

\def\half{{\textstyle{1\over2}}}

\def\bfr{{\bf r}}
\def\bfx{{\bf x}}

\def\p{\partial}
\def\bfr{{\bf r}}

\def\bfv{{\bf v}}

\def\bfw{{\bf w}}

\def\ffrac#1#2{{\textstyle\frac{#1}{#2}}}

\newfont{\caps}{cmcsc10}

\tighten

\begin{document}

\title{Satellite dynamics on the Laplace surface}

\shorttitle{Laplace equilibria}

\shortauthors{TREMAINE, TOUMA, \& NAMOUNI}

\author{Scott Tremaine\altaffilmark{1}, Jihad Touma\altaffilmark{2},
and Fathi Namouni\altaffilmark{3}}
\altaffiltext{1}{School of Natural Sciences, Institute for Advanced
Study, Einstein Drive, Princeton, NJ 08540, USA; tremaine@ias.edu}
\altaffiltext{2}{Department of Physics, American University of Beirut,
PO Box 11--0236, Riad El-Solh, Beirut 1107 2020, Lebanon; 
jihad.touma@gmail.com} 
\altaffiltext{3}{Universit\'e de Nice and CNRS, Observatoire de la C\^ote
  d'Azur, BP 4229, 06304 Nice, France; namouni@obs-nice.fr} 

\begin{abstract}

\noindent
The orbital dynamics of most planetary satellites is governed by the
quadrupole moment from the equatorial bulge of the host planet and the tidal
field from the Sun. On the Laplace surface, the long-term
orbital evolution driven by the combined effects of these forces is zero, so
that orbits have a fixed orientation and shape. The ``classical'' Laplace
surface is defined for circular orbits, and coincides with the planet's
equator at small planetocentric distances and with its orbital plane
at large distances. A dissipative circumplanetary disk should settle to this
surface, and hence satellites formed from such a disk are likely to orbit in
or near the classical Laplace surface.  This paper studies the properties of
Laplace surfaces. Our principal results are: (i) if the planetary obliquity
exceeds $68.875^\circ$ there is a range of semimajor axes in which the
classical Laplace surface is unstable; (ii) at some obliquities and
planetocentric distances there is a distinct Laplace surface consisting of
nested eccentric orbits, which bifurcates from the classical Laplace surface
at the point where instability sets in; (ii) there is also a ``polar'' Laplace
surface perpendicular to the line of nodes of the planetary equator on the
planetary orbit; (iv) for circular orbits, the polar Laplace surface is stable
at small planetocentric distances and unstable at large distances; (v) at the
onset of instability this polar Laplace surface bifurcates into two polar
Laplace surfaces composed of nested eccentric orbits.
\end{abstract}

\section{Introduction}

\noindent
In his study of Jupiter's satellites, \cite{lap05} recognized that the
combined effects of the solar tide and the planet's oblateness induced a ``proper''
inclination in satellite orbits with respect to Jupiter's equator. He remarked
that this proper inclination increases with the distance to the planet, and
defines an orbital plane for circular orbits that lies between the orbital
plane of the planet's motion round the Sun and its equator plane. This plane
is called the Laplace plane.

More generally, the Laplace plane is usually defined as the plane normal to
the axis about which the pole of a satellite's orbit
precesses\footnote{Unfortunately, the term is sometimes also applied to the
  invariable plane, the plane perpendicular to the total angular momentum of
  an $N$-body system.}. The ``Laplace surface'' is the locus of all orbits
that do not precess (i.e., the secular motion of the node and apse vanishes).

In the most common situation, we consider circular satellite orbits around an
oblate planet with non-zero obliquity, traveling around the Sun. The Laplace
surface is then determined by the competition between the interior quadrupole
potential from the equatorial bulge and the external quadrupole potential from
the Sun. Close to the planet, the ``classical'' Laplace surface nearly
coincides with the planetary equator, while at large distances it nearly
coincides with the planetary orbital plane. The transition between these two
orientations occurs near the ``Laplace radius,'' defined below in equation
(\ref{eq:laprad}).

The Laplace surface is important because it traces the shape expected
for a thin gas disk or dissipative particulate ring surrounding the
planet. Thus it is not surprising that many planetary satellites orbit
close to the Laplace surface; those that do probably formed from a
circumplanetary gas disk while those that do not were presumably
either captured from heliocentric orbits or experienced unusual events
in their past history.

The purpose of this paper is to study the properties of the Laplace
surface, including the stability of is generating orbits 
and its generalization to eccentric orbits. Although the Laplace
surface has been known and studied for over two centuries, we believe
that many of the results we present are new. 

\section{Secular equations of motion}

\subsection{The Hamiltonian}

\noindent
The Kepler Hamiltonian that describes the motion of a 
test particle orbiting an isolated point mass $M$ is
\be
H_K=\half v^2-{GM\over r}=-{GM\over 2a};
\ee
here $\bfr$ is the position vector measured from the center of
the planet, $\bfv=\dot\bfr$, $r=|\bfr|$, and $a$ is the semimajor axis of the
test particle. The constants of motion are the Hamiltonian $H_K$ (or
semimajor axis $a$), and the angular momentum and eccentricity vectors 
\be
\bfL=\bfr\times\bfv,\quad \bfe={1\over
  GM}\bfv\times(\bfr\times\bfv) -{\bfr\over r};
\label{eq:cdef}
\ee
these are related to the eccentricity $e$ and semimajor axis of the orbit by
$L^2=GMa(1-e^2)$ and $|\bfe|=e$. 

We now examine how the motion of the test particle is affected by additional
forces from the equatorial bulge of the planet and the Sun.  The
quadrupole potential arising from an oblate planet is
\be
\Phi_p(\bfr)={GMJ_2R_p^2\over r^3}P_2(\cos\theta)=
{GMJ_2R_p^2\over 2r^5}[3(\bfr\cdot\bfn_p)^2-r^2],
\ee
where $J_2$ is the quadrupole gravitational harmonic, $R_p$ is the planetary
radius, $P_2(x)=\half(3x^2-1)$ is a Legendre polynomial, and $\theta$
is the polar angle measured from the planet's spin axis, which is
oriented along the unit vector $\bfn_p$.

The quadrupole potential of the planet may be enhanced by inner satellites.
If the planet hosts $n$ satellites with masses $m_i$, $i=1,\ldots,n$, on
circular orbits in the equatorial plane of the planet with semimajor axes
$a_i$, then at distances $r\gg a_i$ the gravitational potential due to the
satellites can be accounted for by augmenting $J_2$ to $J'_2$, where 
\be
J_2'R_p^2\equiv J_2R_p^2 +\half \sum_{i=1}^n a_i^2m_i/M.  
\label{eq:inner}
\ee 
Values of $J_2$ and $J_2'$ for the giant planets and Pluto are given
in Table \ref{tab:outer}.

\begin{deluxetable}{l|ccccccc}
\tablecaption{Properties of the outer planets\label{tab:outer}}
\startdata
planet  & $a_\odot$ (AU) & $R_p$ (km) & $J_2$    & $J_2'$     & obliquity    &
        $r_L/R_p$ & $r_H/R_p$ \\
\tableline
Jupiter & 5.2029         & 71492 & 0.014696 & 0.045020   & $3.1^\circ$  &
        35.36 & 743.3  \\
Saturn  & 9.5367         & 60330 & 0.016291 & 0.070561   & $26.7^\circ$ &
        48.40 & 1080.1 \\
Uranus  & 19.189         & 26200 & 0.003343 & 0.018699   & $97.9^\circ$ &
        63.96 & 2675.1 \\
Neptune & 30.070         & 25225 & 0.00341  & 0.024069   & $29.6^\circ$ &
        93.20 & 4600.8 \\
Pluto   & 39.482         & 1151  & $-$      & 14.296     & $112.5^\circ$&
        419.6 & 6935.8 
\enddata
\tablecomments{The planet's semimajor axis and radius are
$a_\odot$ and $R_p$. Obliquity is the angle between the planet's spin and
orbital axes, which in the notation of this paper is given by
$\phi_\odot=\cos^{-1}\bfn_p\cdot\bfn_\odot$. The Laplace radius $r_L$ and Hill radius
$r_H$ are defined by
equations (\ref{eq:laprad}) and (\ref{eq:hill}).}
\end{deluxetable}
 
Since the solar tide is assumed to be weak, we may estimate its effects by
averaging over the solar orbital period. We assume that the
planetary orbit has semimajor axis $a_\odot$ and eccentricity $e_\odot$, and denote the
normal to the orbit by the unit vector $\bfn_\odot$. The obliquity of 
the planet is then $\phi_\odot=\cos^{-1}\bfn_\odot\cdot\bfn_p$. Since $r\ll
a_\odot$ we need keep only the quadrupole term in the averaged solar
potential,
\be 
\Phi_\odot(\bfr)={GM_\odot\over 4a_\odot^3(1-e_\odot^2)^{3/2}}
[3(\bfr\cdot\bfn_\odot)^2-r^2].
\ee 
The solar tide also causes the spin axis of the planet to precess, through
its torque on the equatorial bulge. We neglect this effect to keep the
analysis simple, since the precession rate of the planetary spin due to solar
tides is normally much smaller than the precession rate of the satellite
orbit (see \citealt{gol65} and \citealt{bl06} for treatments that include precession of
the planetary spin).

The planet's ``radius of influence'' or ``Hill radius'' is
\be
                 r_H=a_\odot\left(M\over 3M_\odot\right)^{1/3}.
\label{eq:hill}
\ee 
The Hill radius is roughly the point at which $|\Phi_\odot|\sim
GM/r$; beyond the Hill radius the gravitational force experienced by a
satellite is dominated by the solar tide rather than the force from
the planet and most orbits are not bound to the planet.  The Hill
radius also marks the location of the collinear Lagrange points of the
planet \citep{md99}. Our use of the orbit-averaged solar potential
requires that the satellite radius $r\ll r_{\rm H}$.

The total non-Keplerian potential due to the oblate planet, inner
satellites, and Sun is $\Phi=\Phi_p+\Phi_\odot$. 

Now average over the Keplerian orbit of the test particle, which has 
semimajor axis $a$, eccentricity $e$, and orientation specified by the unit
vectors $\bfn$ along the angular momentum vector, $\bfu$ towards pericenter,
and $\bfv=\bfn\times\bfu$. We have
\be
\langle r^2\rangle =a^2(1+\ffrac{3}{2}e^2), \quad
\langle (\bfr\cdot\bfu)^2\rangle=a^2(\half+2e^2), \quad
\langle (\bfr\cdot\bfv)^2\rangle=a^2(\half-\half e^2), 
\label{eq:avg}
\ee
\be
\left\langle {1\over r^3}\right\rangle ={1\over a^3(1-e^2)^{3/2}}, \quad
\left\langle {(\bfr\cdot\bfu)^2\over r^5}\right\rangle =
\left\langle {(\bfr\cdot\bfv)^2\over r^5}\right\rangle =
{1\over 2a^3(1-e^2)^{3/2}}.
\ee
The averaged potential is $\overline\Phi\equiv\overline\Phi_p
+\overline\Phi_\odot$, where 
\begin{eqnarray}
\overline\Phi_p\equiv \langle\Phi_p\rangle & = &
{GMJ_2'R_p^2\over 4a^3(1-e^2)^{3/2}}\left[1-3(\bfn_p\cdot\bfn)^2\right],
\nonumber\\
\overline\Phi_\odot\equiv\langle\Phi_\odot\rangle & = & 
{3GM_\odot a^2\over 4a_\odot^3(1-e_\odot^2)^{3/2}}\left[(\half+2e^2)(\bfn_\odot\cdot\bfu)^2
+(\half-\half e^2)(\bfn_\odot\cdot\bfv)^2-\half e^2 - \ffrac{1}{3}\right].
\end{eqnarray}
The averaged Hamiltonian is then
\be
H=H_K+\overline\Phi_p+\overline\Phi_\odot.
\label{eq:ham}
\ee
By virtue of the orbit averaging, this Hamiltonian is independent of the mean
anomaly, so its conjugate momentum $(GMa)^{1/2}$ is a constant of motion,
which in turn means that the semimajor axis may be treated as 
a constant. 

Now set
\be
\bfj\equiv (1-e^2)^{1/2}\bfn, \qquad \bfe=e\bfu, \qquad \tau=\sqrt{GM\over
  a^3}t, \qquad \epsilon_\odot={M_\odot a^3\over Ma_\odot^3(1-e_\odot^2)^{3/2}}, \qquad
  \epsilon_p={J_2'R_p^2\over a^2}.
\label{eq:www}
\ee 
The angular momentum $\bfL=(GMa)^{1/2}\bfj$ and, as suggested by
the notation, $\bfe$ is the eccentricity vector (eq.\
\ref{eq:cdef}). If we define a dimensionless potential
$\Psi_p+\Psi_\odot=(\overline\Phi_p+\overline\Phi_\odot)a/(GM)$, we have
\begin{eqnarray}
\Psi_p & = &
{\epsilon_p\over 4(1-e^2)^{5/2}}\left[1-e^2-3(\bfj\cdot\bfn_p)^2\right], 
\nonumber \\
\Psi_\odot & = & {3\epsilon_\odot\over8}\left[5(\bfe\cdot\bfn_\odot)^2-
(\bfj\cdot\bfn_\odot)^2-2e^2\right];
\label{eq:psips}
\end{eqnarray}
an unimportant constant has been dropped. 

To repeat, the Hamiltonian $H_K+GM(\Psi_p+\Psi_\odot)/a$ is based on the
assumptions that (i) the precession rate of the planetary spin due to solar
tides is negligible; (ii) the satellite is a massless test particle; (iii) the
Sun is far enough from the planet that the solar tide can be approximated by a
quadrupole; (iv) the satellite is far enough from the planet that the
potential from the planet and the inner satellites can be approximated as a
a monopole plus a quadrupole; (v) the perturbing forces due to
$\Psi_p+\Psi_\odot$ are weak enough that the secular equations of motion can
be used to describe its orbital evolution.

\subsection{Equations of motion}

\noindent
Using equation (\ref{eq:cdef}) it is straightforward to show that the
Poisson brackets of $\bfj$ and $\bfe$ are
\be
\{j_i,j_j\}={1\over\sqrt{GMa}}\epsilon_{ijk}j_k, \quad 
\{e_i,e_j\}={1\over\sqrt{GMa}}\epsilon_{ijk}j_k, \quad
\{j_i,e_j\}={1\over\sqrt{GMa}}\epsilon_{ijk}e_k,
\label{eq:pb}
\ee
where $\epsilon_{ijk}$ is the antisymmetric tensor. 

The time evolution of any variable $f$ determined by the Hamiltonian $H$ is
given by
\be 
   {df\over dt}=\{f,H\}.
\ee
In secular dynamics the semimajor axis is fixed and $H$ can be considered to
be a function only of the shape of the orbit, as expressed by $\bfj$ and
$\bfe$. Then from the chain rule
\be
   {df\over dt}=\{f,\bfj\}\bnabla_\bfj H+\{f,\bfe\}\bnabla_\bfe H,
\ee
where $\bnabla_\bfj$ is the vector $(\p/\p j_1,\p/\p j_2,\p/\p j_3)$ with a
similar definition for $\bnabla_\bfe$. Replacing $f$ successively by $j_i$ and
$e_i$ and using the relations (\ref{eq:pb}), we find
\begin{eqnarray}
   {d\bfj\over dt}&=&-{1\over\sqrt{GMa}}\left(\bfe\times\bnabla_\bfe H
                   +\bfj\times\bnabla_\bfj H\right)  \nonumber \\
   {d\bfe\over dt}&=&-{1\over\sqrt{GMa}}\left(\bfj\times\bnabla_\bfe H
                   +\bfe\times\bnabla_\bfj H\right).
\label{eq:eqmot}
\end{eqnarray}
Since $\bfe$ and $\bfj$ are constants of motion for the Kepler Hamiltonian
$H_K$, the contribution of $H_K$ to the right side of this equation must
vanish, so we can replace $H=H_K+\overline\Phi$ by $\overline\Phi$. Equations
(\ref{eq:eqmot}) date back to \citet[eqs. 204 and 205; see also \citealt{br05}
and references therein]{mil39}.

These equations admit three integrals of motion, $\overline\Phi$,
$\bfj\cdot\bfe$ and $\bfj^2+\bfe^2$. Physically meaningful solutions are
restricted to the four-dimensional manifold on which 
\be
\bfj\cdot\bfe=0,\qquad \bfj^2+\bfe^2=1.
\label{eq:const}
\ee

Replacing $t$ and $\Phi$ by the dimensionless variables $\tau$ and $\Psi$
defined in equation (\ref{eq:www}) we obtain\footnote{There
is a gauge freedom in the definition of $\Psi(\bfj,\bfe)$ since 
the variables $\bfj$ and $\bfe$ are related by (\ref{eq:const}). It is shown
in Appendix \ref{app:a} that the secular equations of motion (\ref{eq:mot}) 
are independent of gauge.}
\begin{eqnarray}
{d\bfj\over d\tau}&=&-\bfj\times\bnabla_\bfj\Psi-\bfe\times\bnabla_\bfe\Psi,
\nonumber \\
{d\bfe\over d\tau}&=&-\bfe\times\bnabla_\bfj\Psi-\bfj\times\bnabla_\bfe\Psi.
\label{eq:mot}
\end{eqnarray}

For the potential given by equations (\ref{eq:psips}) we finally have 
\begin{eqnarray}
{d\bfj\over d\tau}&=&{3\epsilon_\odot\,
\bfj\cdot\bfn_\odot\over 4}\bfj\times\bfn_\odot
-{15\epsilon_\odot\,\bfe\cdot\bfn_\odot\over 4}\bfe\times\bfn_\odot 
+{3\epsilon_p\,\bfj\cdot\bfn_p\over2(1-e^2)^{5/2}}\bfj\times\bfn_p, \nonumber \\
{d\bfe\over d\tau}&=&{3\epsilon_\odot\,\bfj\cdot\bfn_\odot\over 4}
\bfe\times\bfn_\odot
-{15\epsilon_\odot\,\bfe\cdot\bfn_\odot\over 4}\bfj\times\bfn_\odot
+{3\epsilon_p\,\bfj\cdot\bfn_p\over 2(1-e^2)^{5/2}}\bfe\times\bfn_p
\nonumber \\& &\qquad
+\left[\ffrac{3}{2}\epsilon_\odot-\ffrac{3}{4}\epsilon_p{1-e^2-5
(\bfj\cdot\bfn_p)^2\over (1-e^2)^{7/2}}\right]\bfj\times\bfe.
\label{eq:milank}
\end{eqnarray}

To avoid distraction by trivial cases, we shall always assume that
$\epsilon_p>0$ (the planetary quadrupole is non-zero and positive, as expected
for a planet that is oblate or has inner satellites), $\epsilon_\odot>0$
(solar perturbations are non-negligible), and $\bfn_\odot$ is neither
parallel, antiparallel, or perpendicular to $\bfn_p$ (the planetary obliquity
is not $0,\pm\half\pi,\pi$). The plane defined by the planetary spin axis
$\bfn_p$ and the normal to the solar orbit $\bfn_\odot$ will be called the
principal plane.

We shall sometimes use cylindrical coordinates with the $z$-axis oriented
along $\bfn_p\times\bfn_\odot$, so the plane $z=0$ coincides with the
principal plane. The positive $x$-axis is chosen to coincide with
$\bfn_p$. Then $\bfn_\odot$ lies in the $z=0$ plane, so it may be specified by
its azimuthal angle $\phi_\odot$ (the obliquity), which lies in the range
$(0,\pi)$. 

In the limit $\epsilon_p\to0$ equations (\ref{eq:milank}) provide a vector
description of Kozai oscillations, the secular oscillations in the
eccentricity and inclination of (for example) a planet orbiting a member of a
binary star \citep{koz,hol,for}. 

Equations (\ref{eq:milank}) are invariant under the transformations 
\be
\bfe\to-\bfe\quad\hbox{or}\quad
\bfn_\odot\to-\bfn_\odot \quad\hbox{or}\quad \bfn_p\to-\bfn_p\quad\hbox{or}\quad 
(\bfj\to-\bfj,\tau\to-\tau).
\label{eq:symm}
\ee
Invariance under $\bfn_\odot\to-\bfn_\odot$ implies that we can
restrict the range of the obliquity $\phi_\odot$ from $(0,\pi)$ to
$(0,\half\pi)$. 

Recall that equations (\ref{eq:milank}) hold for arbitrary
eccentricity, i.e., they do not represent an expansion that is valid
only for $e\ll1$.

We define the Laplace equilibria to be stationary solutions of equations
(\ref{eq:milank}), and the Laplace surface(s) to be the locus of all orbits 
that are Laplace equilibria. 

\section{Circular Laplace equilibria}

\label{sec:cop}

\noindent
The equations of motion (\ref{eq:milank}) can be solved explicitly in the case
of circular orbits ($\bfe=0$). We banish this discussion to Appendix
\ref{app:b} and focus in this section on the equilibrium solutions, with
$\bfe=0$ and $\bfj=$constant, which we call the circular Laplace equilibria.

In circular Laplace equilibria the second of equations (\ref{eq:milank}) is
satisfied trivially. The first of equations (\ref{eq:milank}) yields
\be
\epsilon_\odot(\bfj\cdot\bfn_\odot)\bfj\times\bfn_\odot
+2\epsilon_p(\bfj\cdot\bfn_p)\bfj\times\bfn_p=0.
\label{eq:eq}
\ee 
Taking the scalar product of this equation with $\bfn_\odot$, and again
with $\bfn_p$, we conclude that either (i)
$\bfj\cdot\bfn_p=\bfj\cdot\bfn_\odot=0$; (ii)
$\bfj\cdot(\bfn_p\times\bfn_\odot)=0$. In the first case $\bfj$ is
perpendicular to the principal plane, and we call this the ``orthogonal'' or
``circular orthogonal'' Laplace equilibrium; in the second case $\bfj$ lies in
the principal plane and we call this the ``coplanar'' or ``circular coplanar''
Laplace equilibrium.

In the coplanar Laplace equilibrium, $\bfj$ lies in the $z=0$ plane, so it may
be specified by its azimuthal angle $\phi$. The equilibrium condition
(\ref{eq:eq}) becomes \be \epsilon_\odot\sin
2(\phi-\phi_\odot)+2\epsilon_p\sin 2\phi=0.
\label{eq:sol}  
\ee 
This equation has four solutions for $\phi$ in a $2\pi$ interval; if
$\phi$ is a solution then $\phi\pm\half\pi$ and $\phi+\pi$ are also solutions.

Equation (\ref{eq:sol}) may also be written as
\be
a^5\sin 2(\phi-\phi_\odot)+2r_L^5\sin 2\phi=0,
\label{eq:sola}  
\ee 
where the Laplace radius $r_L$ is defined by 
\be
r_L^5=J_2'R_p^2a_\odot^3(1-e_\odot^2)^{3/2}{M\over M_\odot}.
\label{eq:laprad}
\ee

\subsection{Stability}

\label{sec:stabc}

\noindent
The stability of the circular Laplace equilibria can be determined by
writing $\bfj=\bfj_0+\bfj_1$ and expanding equations (\ref{eq:milank})
to first order in $\bfj_1$ and $\bfe$:
\begin{eqnarray}
{d\bfj_1\over d\tau}&=&{3\epsilon_\odot
\bfj_1\cdot\bfn_\odot\over 4}\bfj_0\times\bfn_\odot+{3\epsilon_\odot
\bfj_0\cdot\bfn_\odot\over 4}\bfj_1\times\bfn_\odot
+{3\epsilon_p\bfj_1\cdot\bfn_p\over2}\bfj_0\times\bfn_p
+{3\epsilon_p\bfj_0\cdot\bfn_p\over2}\bfj_1\times\bfn_p, \nonumber \\
{d\bfe\over d\tau}&=&{3\epsilon_\odot\bfj_0\cdot\bfn_\odot\over 4}
\bfe\times\bfn_\odot
-{15\epsilon_\odot\bfe\cdot\bfn_\odot\over 4}\bfj_0\times\bfn_\odot
+{3\epsilon_p\bfj_0\cdot\bfn_p\over 2}\bfe\times\bfn_p
\nonumber \\& &\qquad\qquad\qquad\qquad\qquad
+\left\{\ffrac{3}{2}\epsilon_\odot-\ffrac{3}{4}\epsilon_p[1-5
(\bfj_0\cdot\bfn_p)^2]\right\}\bfj_0\times\bfe.
\label{eq:milanklin}
\end{eqnarray}
Note that the two equations are decoupled: the linearized evolution of the
orientation of the orbital plane, specified by $\bfj$, is independent of the
linearized evolution of the eccentricity and apse direction, specified by
$\bfe$.

\begin{figure}
\plotone{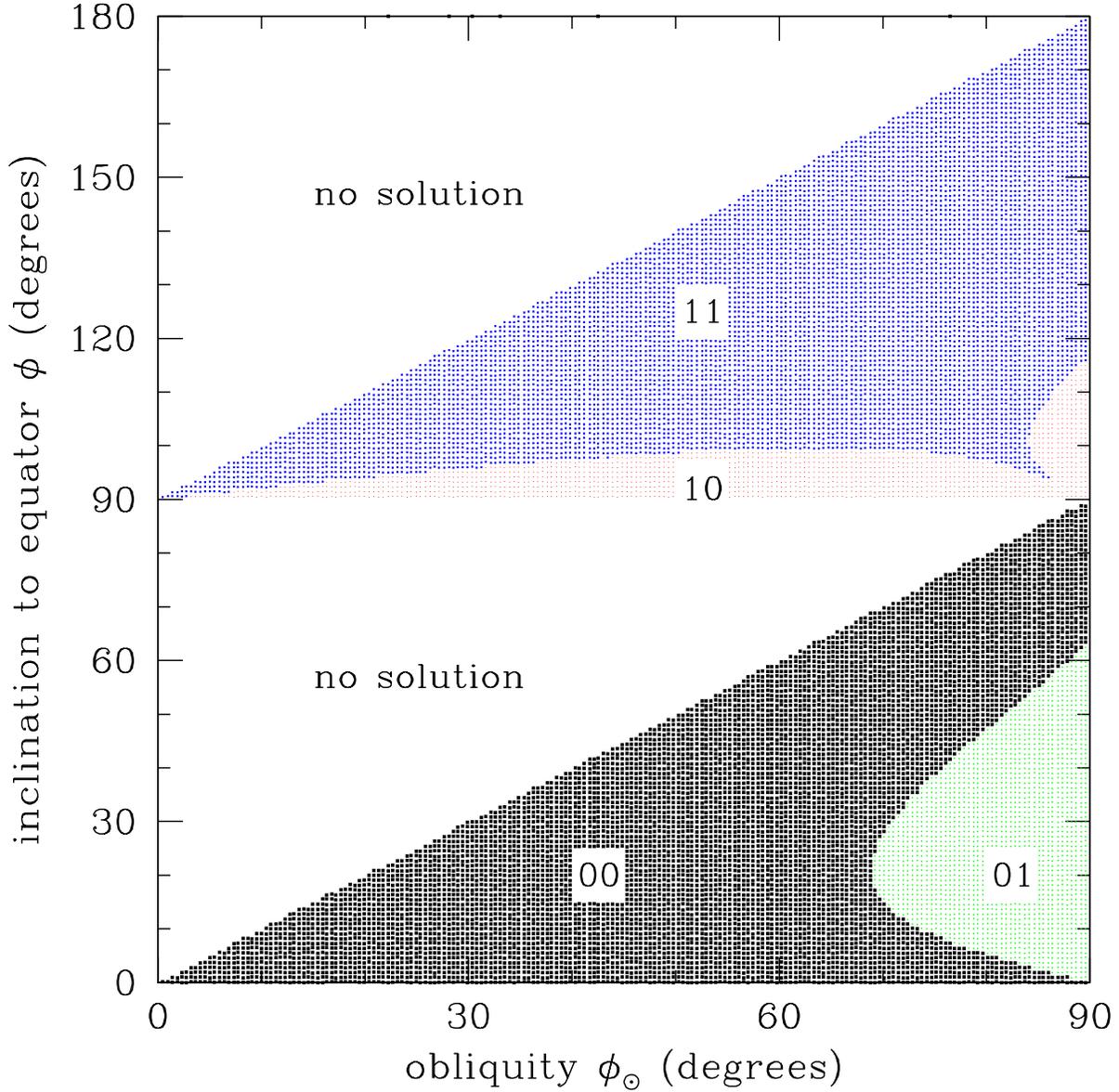}
\caption{The regions in which stable, circular, coplanar Laplace equilibria exist are
  marked by heavy black stippling and the label ``00''. In the more lightly
  stippled regions, circular coplanar equilibria exist but are unstable, according to
  equations (\ref{eq:stabcop}) and (\ref{eq:stabcope}). The first figure in
  each label is ``0'' or ``1'' according to whether the solution is stable or
  unstable to changes in the orbit plane orientation, described by $\bfj$.
  The second figure is ``0'' or ``1'' according to whether the solution is
  stable or unstable to changes in the eccentricity, described by $\bfe$.  The
  vertical coordinate is the inclination of the orbit relative to the
  planetary equator, $\phi$. The horizontal coordinate is the obliquity of the
  planet relative to the ecliptic, $\phi_\odot$. The results are unchanged if
  $\phi\to\phi+\pi$ or $\phi_\odot\to\phi_\odot+\pi$ or
  $(\phi_\odot,\phi)\to(\pi-\phi_\odot,\pi-\phi)$.}
\label{fig:stab}
\end{figure} 

It is easy to show using the equilibrium equation (\ref{eq:eq}) that a trivial
solution of the first of these equations is $\bfj_1=k\bfj_0$ where $k\ll1$ is
a constant. In other words, the eigenvalue equation in $\lambda$ obtained by
assuming $\bfj_1\propto\exp(\lambda t)$ always has one zero eigenvalue. This
is an unphysical solution of the linearized equations since the constant of
motion $\bfj^2+\bfe^2=1$ requires that $\bfj_0\cdot\bfj_1=0$.

\subsubsection{The circular orthogonal Laplace equilibrium}

\label{sec:orthoc}

\noindent
In the orthogonal equilibrium equations (\ref{eq:milanklin}) simplify to
\begin{eqnarray}
{d\bfj_1\over d\tau}&=&{3\epsilon_\odot
\bfj_1\cdot\bfn_\odot\over 4}\bfj_0\times\bfn_\odot
+{3\epsilon_p\bfj_1\cdot\bfn_p\over2}\bfj_0\times\bfn_p, \nonumber \\
{d\bfe\over d\tau}&=&
-{15\epsilon_\odot\bfe\cdot\bfn_\odot\over 4}\bfj_0\times\bfn_\odot
+\left(\ffrac{3}{2}\epsilon_\odot-\ffrac{3}{4}\epsilon_p\right)\bfj_0\times\bfe.
\label{eq:milanklini}
\end{eqnarray}
These can be converted to eigenvalue equations by assuming
$\bfj_1,\bfe_1\propto \exp(\lambda t)$. 
To analyze the first equation, set $u=\bfj_1\cdot\bfn_\odot$,
$v=\bfj_1\cdot\bfn_p$. Taking the scalar product of $d\bfj_1/dt$ with
$\bfn_\odot$ and $\bfn_p$ we find
\be
{du\over d\tau}=\ffrac{3}{2}\epsilon_p\bfn_\odot\cdot(\bfj_0\times\bfn_p) v\quad;
\quad 
{dv\over d\tau}=\ffrac{3}{4}\epsilon_\odot\bfn_p\cdot(\bfj_0\times\bfn_\odot) u.
\ee
these can be combined to show that either (i) $u=v=0$, $\lambda=0$; this
is the unphysical solution noted above; or (ii) 
\be
 \lambda^2=-\ffrac{9}{8}\epsilon_p\epsilon_\odot[
\bfj_0\cdot(\bfn_\odot\times\bfn_p)]^2.
\ee
since $\lambda^2<0$ in this case, $\bfj_1$ is oscillatory, so the circular orthogonal
equilibrium is linearly stable to variations in the angular momentum vector
$\bfj$. The quantity in square brackets is just
$\sin\phi_\odot$, where $\phi_\odot$ is the obliquity.

The second of equations (\ref{eq:milanklini}) can be analyzed similarly. A
trivial solution is $\bfe\propto \bfj_0$, $\lambda=0$; this solution is
unphysical since the constant of motion $\bfj\cdot\bfe=0$ requires
$\bfj_0\cdot\bfe=0$. The other eigenvalues are given by 
\be
\lambda^2=\ffrac{9}{16}(3\epsilon_\odot+\epsilon_p)(2\epsilon_\odot-\epsilon_p).
\ee 
Thus the circular orthogonal equilibrium is stable to variations in eccentricity
$\bfe$ if and only if 
\be 
2\epsilon_\odot<\epsilon_p\quad\hbox{or}\quad a < 2^{-1/5} r_L.  
\label{eq:al}
\ee

\subsubsection{The circular coplanar Laplace equilibrium}

\label{sec:unscirc}

\noindent
In the circular coplanar equilibrium, the first of the linearized equations
(\ref{eq:milanklin}) implies that $\bfj_1\propto \exp(\lambda t)$ where
\be
\lambda^2= -\ffrac{9}{4}\epsilon_p^2\cos^2\phi
             -\ffrac{9}{16}\epsilon_\odot^2\cos^2(\phi-\phi_\odot)
          -\ffrac{9}{16}\epsilon_p\epsilon_\odot[\cos2\phi
             +\cos2(\phi-\phi_\odot)+2\cos2\phi_\odot].
\ee
Substituting the equilibrium condition (\ref{eq:sol}) we have
\be
\lambda^2=-{9\epsilon_\odot^2\over 16}{\sin^2\phi_\odot
           \cos\phi_\odot\over\sin^2\phi\cos\phi}
           \cos(\phi-\phi_\odot).
\label{eq:stabcop}
\ee 
The regions in which $\lambda^2>0$ (instability) are marked by red and blue
stippling and the labels ``10'' and ``11'' in Figure \ref{fig:stab}.
All equilibria with $0<\phi<\half\pi$ are stable to perturbations of this
kind (variations in $\bfj$ but not $\bfe$) and 
equilibria with $\half\pi<\phi<\pi$ are unstable. 

The circular coplanar equilibria that are stable in this sense have
$\phi\to 0$ as $a\to 0$ and $\phi\to\phi_\odot$ as $a\to\infty$, so
the corresponding Laplace surface coincides with the equator of the
planet at small radii and with the planet's orbital plane at large
radii. We call this the ``classical'' Laplace surface since this was
the surface discovered by Laplace and the one that has been the focus
of most work on this subject.

The second of the linearized equations (\ref{eq:milanklin}) implies that
$\bfe\propto \exp(\lambda t)$ where 
\begin{eqnarray}
&&\lambda^2=
-{9\epsilon_p^2\over 16}[5\cos^4\phi-2\cos^2\phi+1]
+{9\epsilon_\odot^2\over 16}[6-7\cos^2(\phi-\phi_\odot)] \\
&& +{9\epsilon_p\epsilon_\odot\over
      64}[5-6\cos^2\phi-6\cos^2(\phi-\phi_\odot)
    -6\sin2\phi\sin2(\phi-\phi_\odot)-9\cos2\phi\cos2(\phi-\phi_\odot)]. \nonumber
\end{eqnarray}
Substituting the equilibrium solution (\ref{eq:sol}) we have
\begin{eqnarray}
\lambda^2&=&-{9\epsilon_\odot^2\over 2048\sin^22\phi}\bigg\{-106 +
  24\cos2\phi+146\cos4\phi -100\cos(6\phi-2\phi_\odot)\nonumber \\
  &&-24\cos(2\phi-4\phi_\odot)+224\cos(2\phi-2\phi_\odot)-54\cos(4\phi-4\phi_\odot)
 -8\cos2\phi_\odot-11\cos4\phi_\odot\nonumber \\
  && -124\cos(2\phi+2\phi_\odot)
  +25\cos(8\phi-4\phi_\odot) +8\cos(4\phi-2\phi_\odot)\bigg\}.  
\label{eq:stabcope}
\end{eqnarray}
The regions in which $\lambda^2>0$ (instability) are marked by green and blue 
stippling and the labels ``01'' and ``11'' in Figure \ref{fig:stab}.

\begin{figure}
\plotone{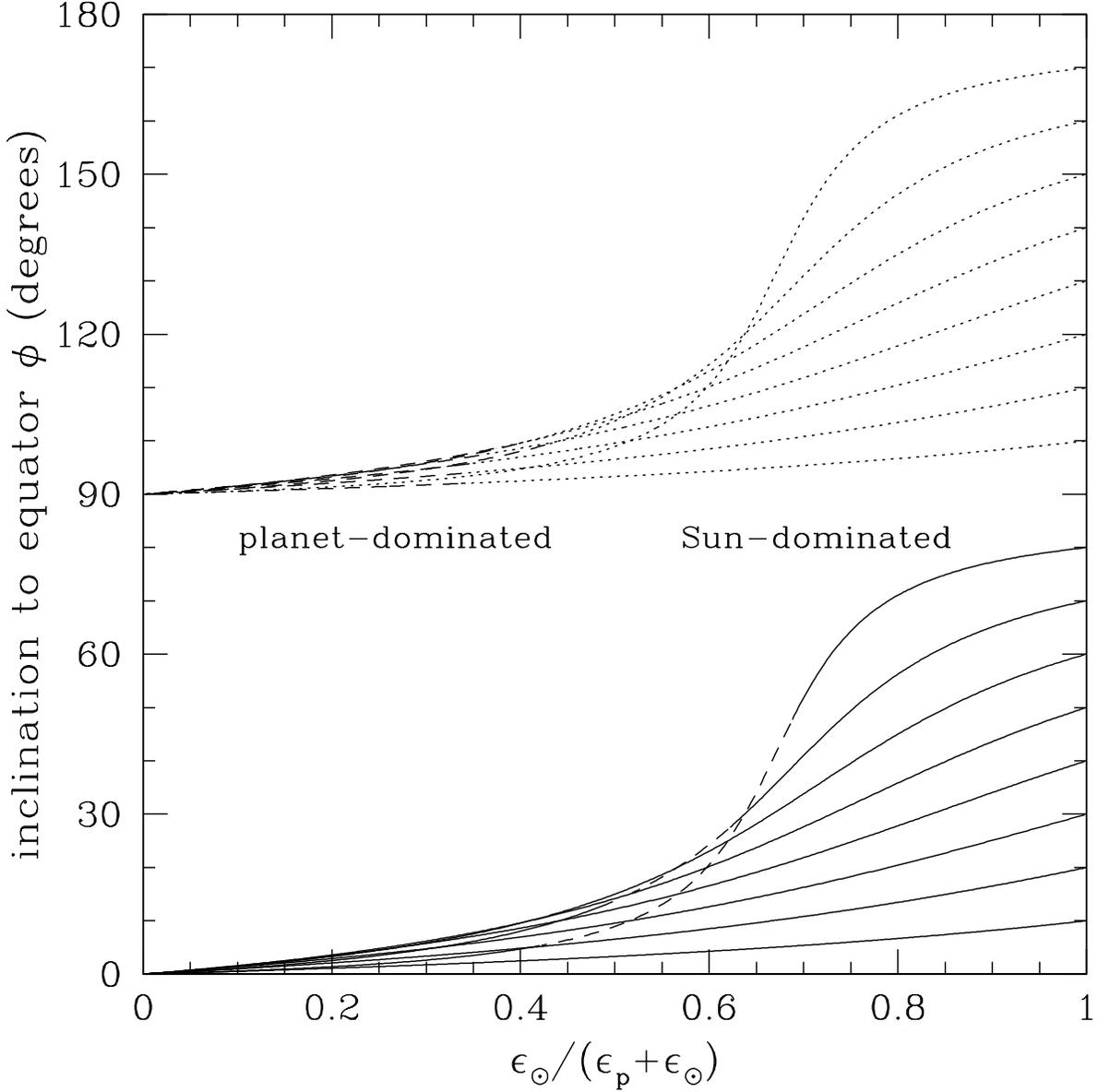}
\caption{Circular coplanar Laplace equilibria. The vertical axis is the angle $\phi$
  between the pole of the planet and the orbital pole of the satellite (the
  inclination of the orbit relative to the planet's equator). Solutions are
  shown for eight values of the planetary obliquity, $\phi_\odot=
  10\deg,\ldots,80\deg$. There are two equilibrium curves for each obliquity,
  one with $0<\phi<90\deg$ and the other with $90\deg<\phi<180\deg$. Solid
  lines denote stable equilibria, while dashed and dotted lines denote
  equilibria with one or two unstable roots respectively. All equilibria with
  $\phi>90\deg$ are unstable. The classical equilibria are those with
  $\phi<90^\circ$. The horizontal axis represents the relative
  strength of perturbations from the solar tide and the planetary quadrupole
  (eq.~\ref{eq:www}); the planetary quadrupole dominates on the left side of
  the figure and the solar quadrupole on the right.}
\label{fig:cop}
\end{figure}

The inclination to the planetary equator $\phi$ is plotted as a function of
the strength of the planetary quadrupole $\epsilon_p$ in Figure \ref{fig:cop},
for obliquities $\phi_\odot=10\deg,20\deg,\ldots,80\deg$.  Solutions
for $\phi_\odot>90\deg$ can be obtained by the transformation
$(\phi,\phi_\odot)\to(\pi-\phi,\pi-\phi_\odot)$. 

\begin{figure}
\plotone{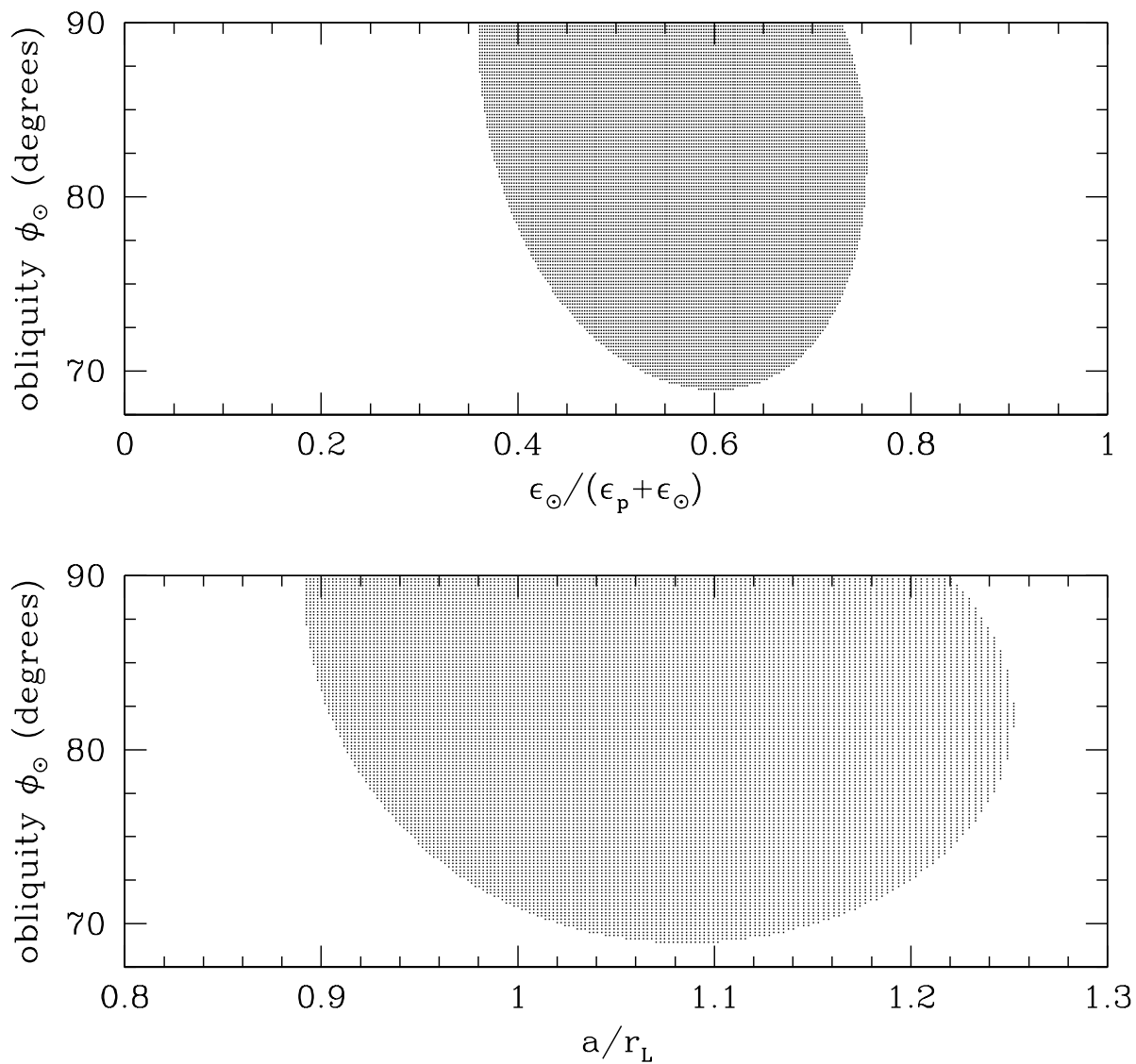}
\caption{Regions in which the classical Laplace equilibria are unstable are
  stippled. The top panel shows the unstable range of obliquities as a
  function of the ratio of the solar and planetary perturbation strengths
  (eq.~\ref{eq:www}) and the bottom panel shows the unstable obliquities as a
  function of semimajor axis (in units of the Laplace radius $r_L$, eq.\
  \ref{eq:laprad}). All instabilities are in the eccentricity vector; the
  angular-momentum vector is stable.}
\label{fig:unstable}
\end{figure} 

\begin{figure}
\plotone{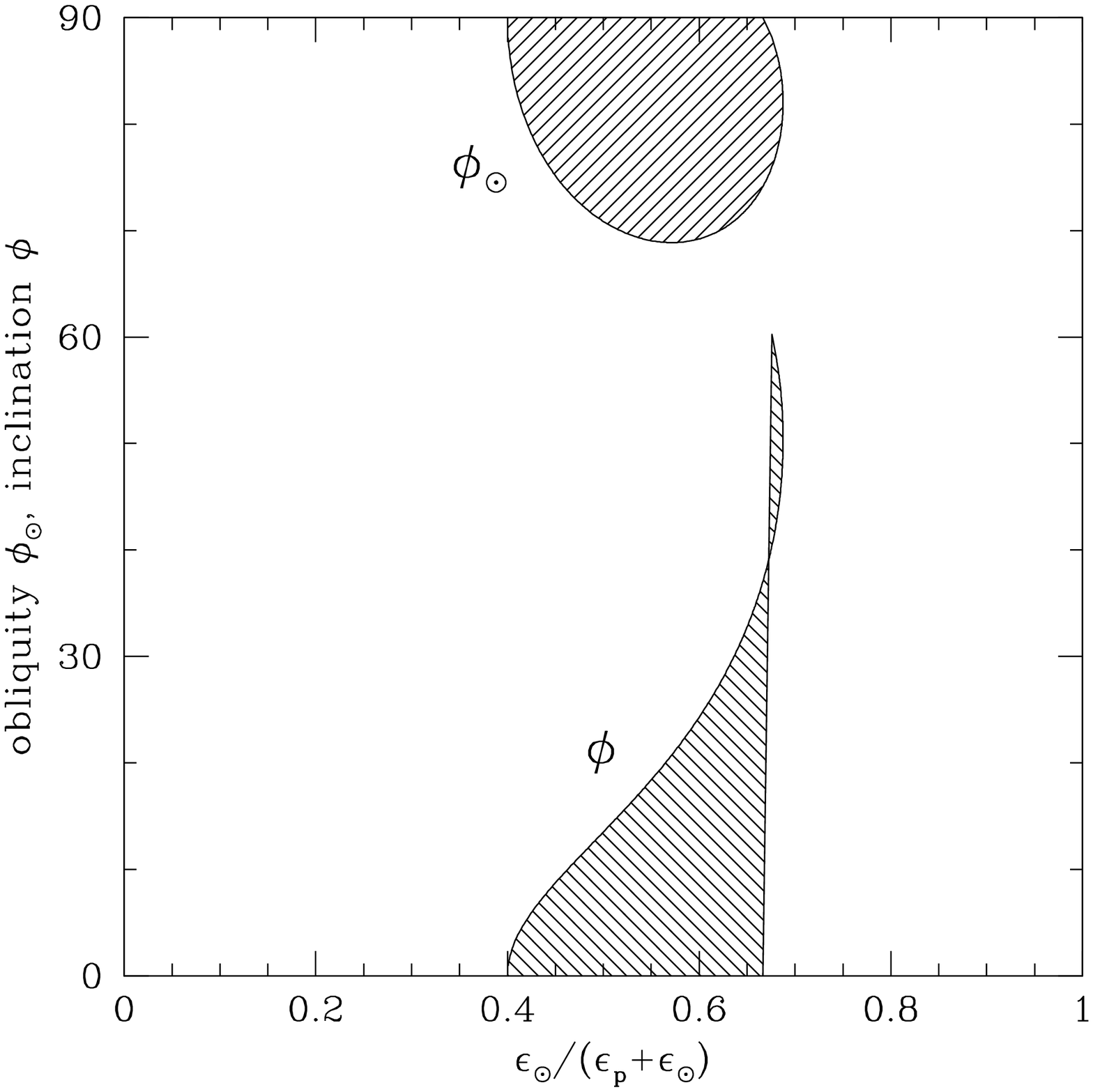}
\caption{Regions in which the classical Laplace equilibria are unstable are
  shaded.} 
\label{fig:kerr5}
\end{figure} 

Unstable equilibria are shown in Figure \ref{fig:cop} by dotted or dashed
lines. All equilibria with $90\deg<\phi<180\deg$ are unstable. In addition,
some of the classical equilibria ($\phi<90\deg$) are unstable to
eccentricity growth. These are shown as stippled or shaded regions in Figures
\ref{fig:unstable} and \ref{fig:kerr5}. Instability first appears at obliquity
$\phi_\odot=68.875\deg$ and is restricted to semimajor axes between about 0.9
and 1.25 times the Laplace radius $r_L$ (eq.\ \ref{eq:laprad}).

\section{Eccentric Laplace equilibria}
\label{sec:circ}

\noindent
We now look for stationary solutions to equations (\ref{eq:milank}) in
which the eccentricity $e=|\bfe|$ is non-zero (recall that
$|\bfj|=(1-e^2)^{1/2}$).  It can be shown (see Appendix \ref{app:c})
that all such solutions either have both $\bfj$ and $\bfe$ in the
principal plane defined by $\bfn_\odot$ and $\bfn_p$ (the
``coplanar-coplanar'' or ``eccentric coplanar-coplanar'' solution), or
one of $\bfj$ and $\bfe$ in the principal plane and the other
orthogonal to it (the ``coplanar-orthogonal'' or
``orthogonal-coplanar'' equilibrium if $\bfj$ or $\bfe$, respectively,
lies in the principal plane).

\subsection{Eccentric coplanar-coplanar Laplace equilibrium}
\label{sec:coco}

\noindent
In this case $\bfj$ and $\bfe$ lie in the principal plane, 
and equations (\ref{eq:milank}) with $d\bfj/d\tau=d\bfe/d\tau=0$ yield
\begin{eqnarray}
2\epsilon_p\sin 2\phi &=& \epsilon_\odot(1-e^2)^{3/2}(1+4e^2) 
\sin 2(\phi_\odot-\phi) \nonumber \\
{}\epsilon_p[1-3\cos^2\phi] &=&
\epsilon_\odot(1-e^2)^{5/2}[1-4\sin^2(\phi_\odot-\phi)].
\end{eqnarray}
These can be used to solve for $e^2$ and $\phi$, given
$\phi_\odot$ and $\epsilon_p/\epsilon_\odot$.  The solutions are physical only
if $0<e^2<1$.

\begin{figure}
\plotone{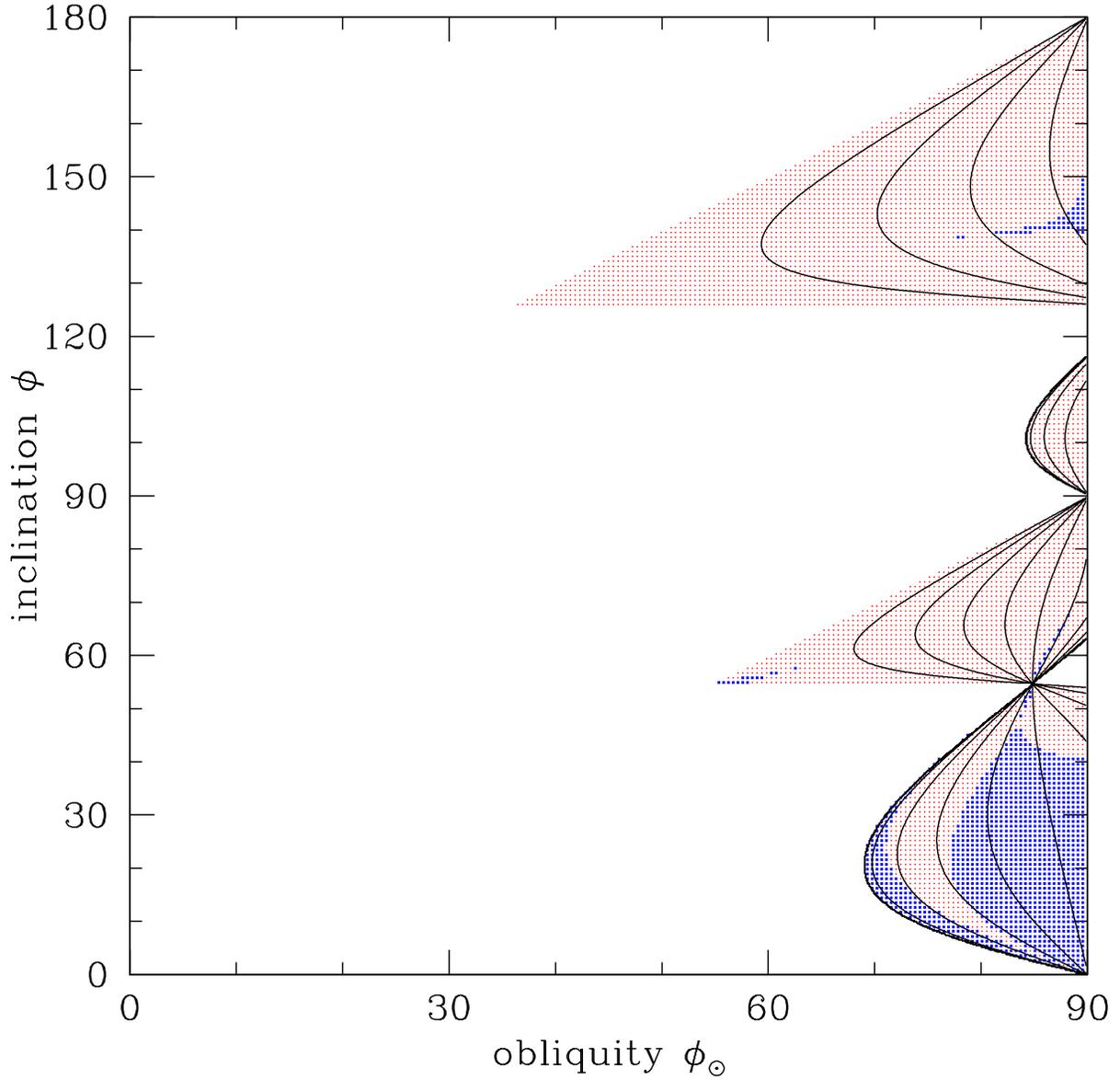}
\caption{Regions in which eccentric coplanar-coplanar Laplace equilibria exist are
  stippled. Heavy blue and light red stippling mark regions in which the
  equilibria are respectively stable and unstable. Contours mark
  eccentricities of $0.1,0.2,\ldots,0.9$. Compare to Figure \ref{fig:stab}.}
\label{fig:stablecc1}
\end{figure} 

The regions in $\phi_\odot\hbox{--}\phi$ space in which eccentric coplanar-coplanar
Laplace equilibria exist are shown in Figure \ref{fig:stablecc1} by
stipples. These regions are bounded, in part, by the lines $\phi=\phi_\odot$,
$\phi=\phi_\odot+\half\pi$, and
$\phi=\pm\cos^{-1}(1/\surd{3})=54.7\deg,125.3\deg$.  Heavy stippling marks
regions in which the equilibria are stable.

\begin{figure}
\plotone{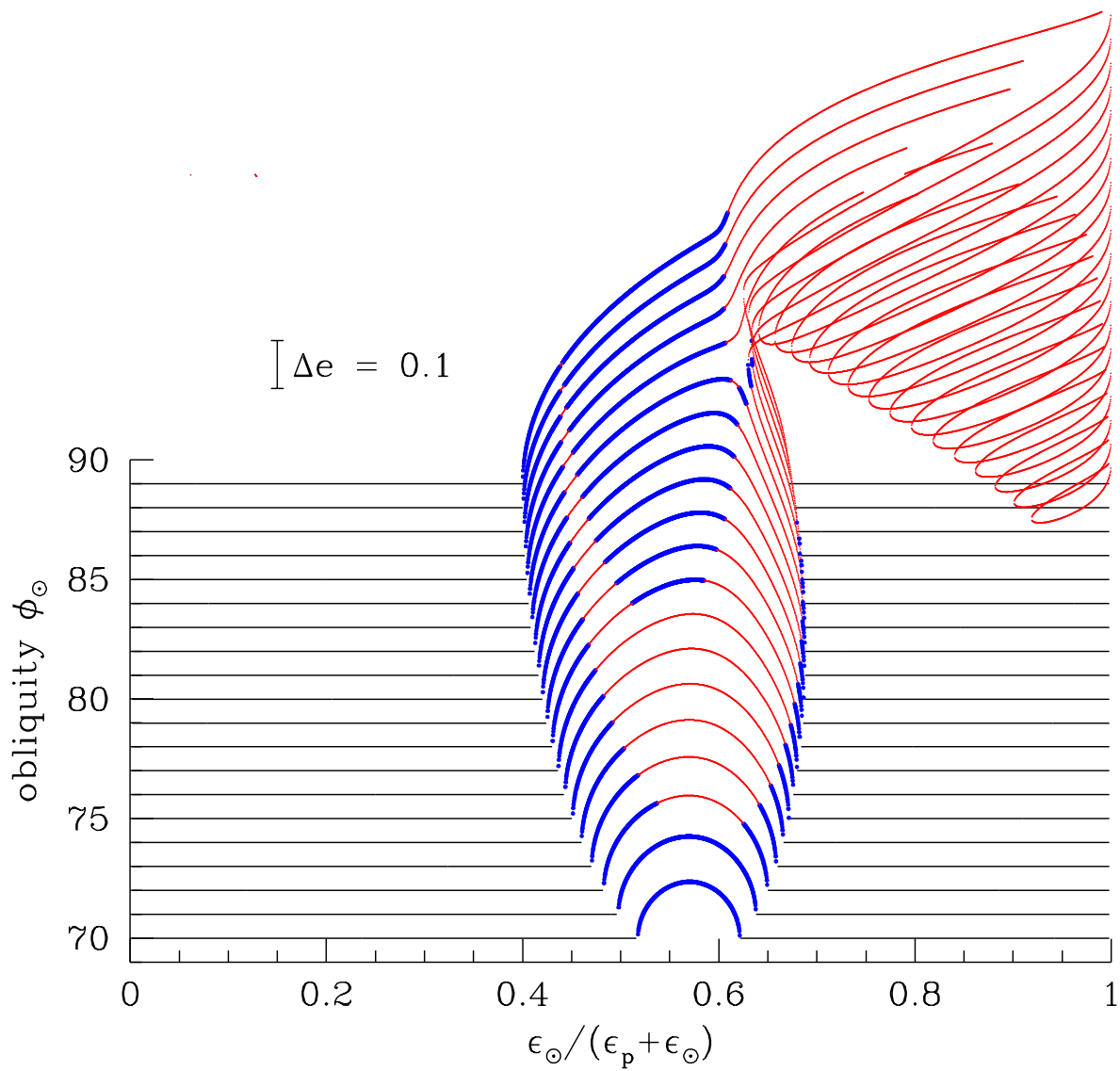}
\caption{The relation of the circular coplanar Laplace equilibria to the
  eccentric coplanar-coplanar equilibria. The gaps in the solid horizontal
  lines represent regions in which the circular coplanar equilibria are
  unstable. The curved lines represent the eccentric coplanar-coplanar Laplace
  equilibria with inclination $\phi<90\deg$. The $y$-coordinate of these lines
  is $\phi_\odot+20e$ where $e$ is the eccentricity of the solution; the
  curves are heavy blue or light red according to whether the solution is
  stable or unstable. The figure shows that the eccentric coplanar-coplanar
  equilibria bifurcate from the circular coplanar equilibria.}
\label{fig:coco}
\end{figure}

The stable regions are small for $\phi>54.7\deg$ so we will focus on the
region $\phi<54.7\deg$. The eccentric coplanar-coplanar equilibria in this
region are closely related to the circular coplanar equilibria discussed in
\S\ref{sec:cop}. We found there that the circular coplanar Laplace equilibria
were unstable for a range of $\epsilon_p/\epsilon_\odot$ when the obliquity
$\phi_\odot>68.875\deg$.  Figure \ref{fig:coco} shows the range in which the
circular coplanar equilibria are stable for
$\phi_\odot=70\deg,71\deg,\ldots,89\deg$ as solid horizontal lines, with gaps
marking the unstable range. Superimposed on these lines are the eccentric
coplanar-coplanar equilibria, marked by heavy blue or light red curves
depending on whether they are stable or unstable. The height of these curves
is proportional to the eccentricity of the equilibrium. The figure shows that
the eccentric coplanar-coplanar equilibria bifurcate from the circular
coplanar equilibrium at the point where the circular coplanar equilibrium
becomes unstable. The structure of the secular Hamiltonian at the bifurcation
is that of the standard resonant Hamiltonian at $j+2:j$ orbital resonances ($k=2$ in
the notation of \citealt{bg84}). 

For $68.875\deg < \phi_\odot < 71.072\deg$ the eccentric
coplanar-coplanar solution exists in a limited range of
$\epsilon_p/\epsilon_\odot$ and is stable throughout this region. For
$\phi_\odot > 71.072\deg$ the eccentric coplanar-coplanar solution is
unstable for some part of the range of $\epsilon_p/\epsilon_\odot$ in
which it exists. Let us imagine a satellite in the circular coplanar
Laplace equilibrium at (say)
$\epsilon_\odot/(\epsilon_p+\epsilon_\odot)=0.8$. If
$\epsilon_\odot/(\epsilon_p+\epsilon_\odot)$ slowly decreases (for
example, because the satellite is slowly migrating toward the planet)
then we expect that (i) for $\phi_\odot < 68.875\deg$ the satellite
will always remain in a circular coplanar Laplace equilibrium; (ii)
for $68.875\deg < \phi_\odot < 71.072\deg$ the satellite will transfer
onto the eccentric coplanar-coplanar equilibrium, its eccentricity
will then grow as its spirals in, reach a maximum depending on
$\phi_\odot$, then shrink back to zero, at which point it will rejoin
the sequence of circular coplanar Laplace equilibria and spiral into
the planet on a circular orbit; (iii) for $\phi_\odot > 71.072\deg$
the satellite will transfer onto the eccentric coplanar-coplanar
equilibrium, and its eccentricity will grow until the equilibrium
orbit becomes unstable, at which point it presumably undergoes large
oscillations in eccentricity and inclination. Numerical simulations of
this evolution are described in \S\ref{sec:num}; we shall find that
(i) is correct, but that (ii) and (iii) need to be qualified in that
orbits track the sequence of eccentric equilibria less well than they
imply.

An interesting but difficult question, which we do not attempt to
answer here, is the behavior of a dissipative gas or particulate disk
in the region where the classical Laplace surface is unstable. There
are at least three alternatives: (i) The dissipative forces suppress
the secular instabilities in the circular coplanar Laplace equilibria,
thereby allowing the disk to occupy the classical Laplace
surface. (ii) The disk occupies the surface defined by the eccentric
coplanar-coplanar equilibria. A possible problem is the large
eccentricity gradients in this sequence of equilibria: a flat, cold
disk composed of aligned eccentric orbits cannot exist if
$|a(de/da)|>1$, because the orbits cross. This condition does not
apply directly to the Laplace surface because it is not flat;
nevertheless, the large eccentricity gradients may lead to strong
shears that destabilize the disk. (iii) A gap opens in the disk where
the classical Laplace surface is unstable.

\subsection{Eccentric coplanar-orthogonal Laplace equilibrium}

\noindent
In this case $\bfj$ lies in the principal plane and $\bfe$ is normal to it, so
$$
   \bfn_\odot\cdot(\bfj\times\bfn_p)=0,\quad \bfe\cdot\bfn_\odot=0, \quad
   \bfe\cdot\bfn_p=0. 
$$ 
The first of equations (\ref{eq:milank}) with $d\bfj/d\tau=0$
yields 
\be 
\epsilon_\odot(1-e^2)^{5/2} \sin 2(\phi-\phi_\odot)+2\epsilon_p\sin 2\phi=0.
\label{eq:coorth}
\ee 
We introduce a vector $\bfw$ defined by $\bfw\times\bfj=\bfe$ or
$\bfw=\bfj\times\bfe/(1-e^2)$; $\bfw$ lies in the principal plane and
$\bfw\cdot\bfj=\bfw\cdot\bfe=0$.  Substituting for $\bfe$ in the second of
equations 
(\ref{eq:milank}) with $d\bfe/d\tau=0$ and $\bfe\cdot\bfn_\odot=0$ yields
\begin{eqnarray}
0 &=& \epsilon_\odot\,\bfj\cdot\bfn_\odot[\bfj(\bfw\cdot\bfn_\odot) - 
\bfw(\bfj\cdot\bfn_\odot)] +{2\epsilon_p\,\bfj\cdot\bfn_p\over(1-e^2)^{5/2}}
[\bfj(\bfw\cdot\bfn_p)-\bfw(\bfj\cdot\bfn_p)] \nonumber \\
&&\qquad\qquad +\left[2\epsilon_\odot
-\epsilon_p{1-e^2-5(\bfj\cdot\bfn_p)^2\over(1-e^2)^{7/2}}\right]j^2\bfw.
\end{eqnarray}
The vectors $\bfj$ and $\bfw$ are linearly independent, so the coefficient of
each must be zero. The coefficient of $\bfj$ vanishes if equation
(\ref{eq:coorth}) is satisfied. The condition that the coefficient of $\bfw$
vanishes is 
\be
\epsilon_\odot[2-\cos^2(\phi_\odot-\phi)]
+{\epsilon_p\over(1-e^2)^{5/2}}(3\cos^2\phi-1)=0. 
\label{eq:coortha}
\ee

\begin{figure}
\includegraphics[clip=true,width=\hsize]{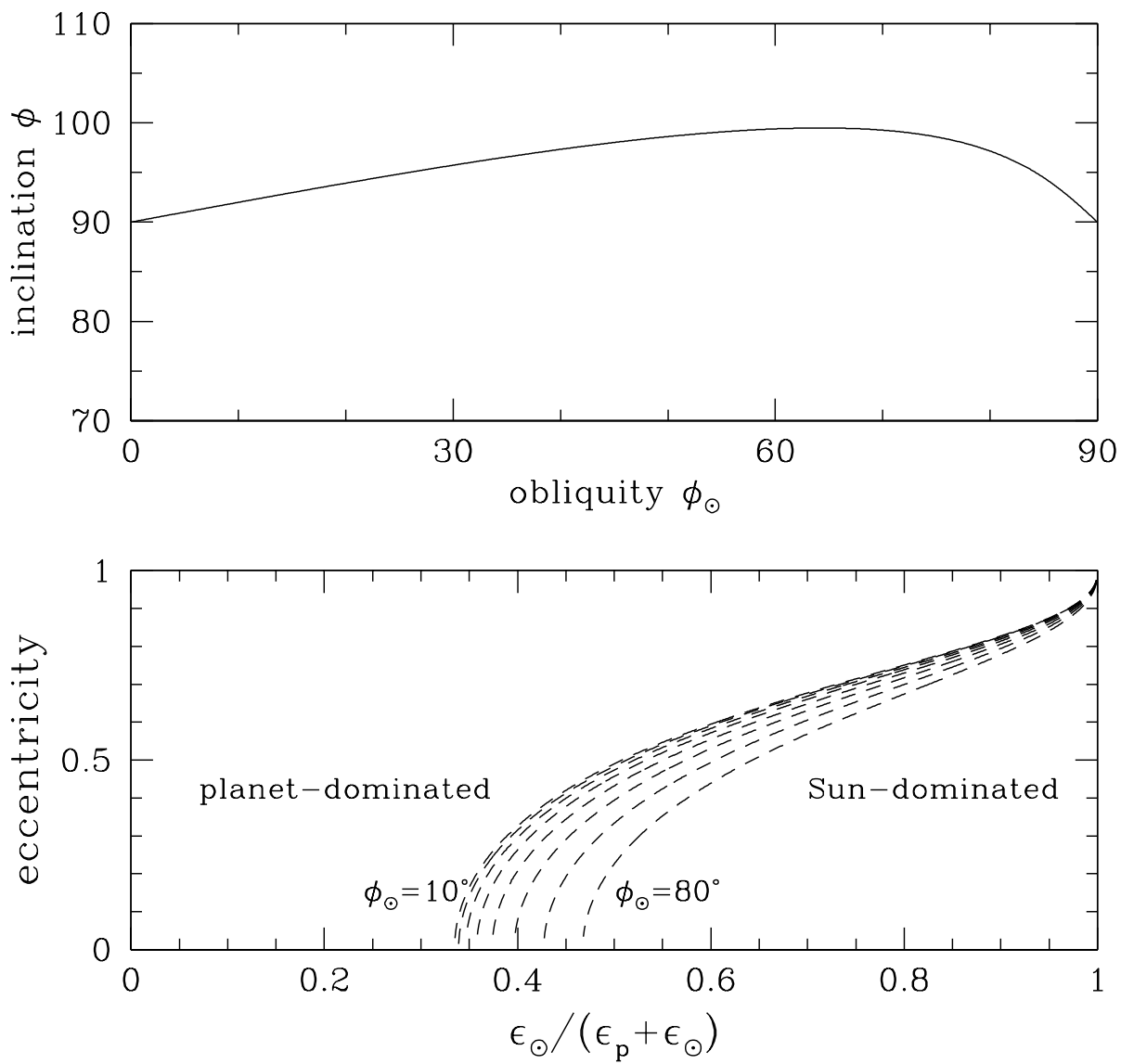}
\caption{(Top) Inclination $\phi$ as a function of the obliquity
  $\phi_\odot$ for the eccentric coplanar-orthogonal Laplace equilibrium, as obtained by
  solving equation (\ref{eq:wwpp}). (Bottom) Eccentricity of the
  coplanar-orthogonal equilibrium for obliquity
  $\phi_\odot=10,20,\ldots,80\deg$ (left to right). The dashed lines indicate
  that all of these equilibria have one unstable mode.}
\label{fig:stableco}
\end{figure}

Equations (\ref{eq:coorth}) and (\ref{eq:coortha}) can be combined to
eliminate $e$, $\epsilon_\odot$, and $\epsilon_p$:
\be 
2[2-\cos^2(\phi_\odot-\phi)]\sin 2\phi
+(3\cos^2\phi-1)\sin 2(\phi_\odot-\phi)=0. 
\label{eq:wwpp}
\ee 
This result determines the inclination $\phi$ between 
the satellite orbit and the planetary spin in terms of the
obliquity $\phi_\odot$, independent of $\epsilon_p$, $\epsilon_\odot$, or
$e$. Numerical solution of this equation for $0\le\phi_\odot\le\half\pi$ shows
that there are two solutions for each value of $\phi_\odot$, but the upper
solution is unphysical, because these values of $\phi$ and $\phi_\odot$ yield
$\epsilon_\odot(1-e^2)^{5/2}/\epsilon_p<0$ in equation (\ref{eq:coorth}). Thus
there is a unique inclination $\phi$ for each value of the obliquity 
$\phi_\odot$, as shown in the top panel of Figure \ref{fig:stableco}. 

The bottom panel of Figure \ref{fig:stableco} shows the eccentricity of the
coplanar-orthogonal Laplace equilibria. All of these equilibria are unstable. 

\subsection{Eccentric orthogonal-coplanar Laplace equilibrium}

\label{sec:orthco}

\noindent
In this case $\bfe$ lies in the principal plane and $\bfj$ is normal to it, so
\be
\bfj\cdot\bfn_p=\bfj\cdot\bfn_\odot=0, \quad 
\bfe\cdot(\bfn_p\times\bfn_\odot)=0.
\ee
The solution of equations (\ref{eq:milank}) with $d\bfj/d\tau=d\bfe/d\tau=0$
requires that 
\be 
\bfe\cdot\bfn_\odot=0, \quad 2\epsilon_\odot(1-e^2)^{5/2}=\epsilon_p.  
\ee 
Solutions exist whenever $2\epsilon_\odot>\epsilon_p$.

Linear stability analysis shows that small perturbations in $\bfj$ or $\bfe$
grow as $\exp(\lambda t)$ where
\be
\lambda^2={9\epsilon_\odot^{8/5}\epsilon_p^{2/5}\over
  2^{12/5}}\sin^2\phi_\odot\quad\hbox{or}\quad 
\lambda^2={225\epsilon_\odot\over
  8}\left[1-(\epsilon_p/2\epsilon_\odot)^{2/5}\right];
\ee
since $\lambda^2<0$ the eccentric orthogonal-coplanar solutions are stable. 

Comparison with the results of \S\ref{sec:orthoc} shows that these
(eccentric) equilibria bifurcate from the (circular) orthogonal
equilibrium sequence at the semimajor axis $a=2^{-1/5}r_L$ where the
latter becomes unstable.

\begin{figure}
\plotone{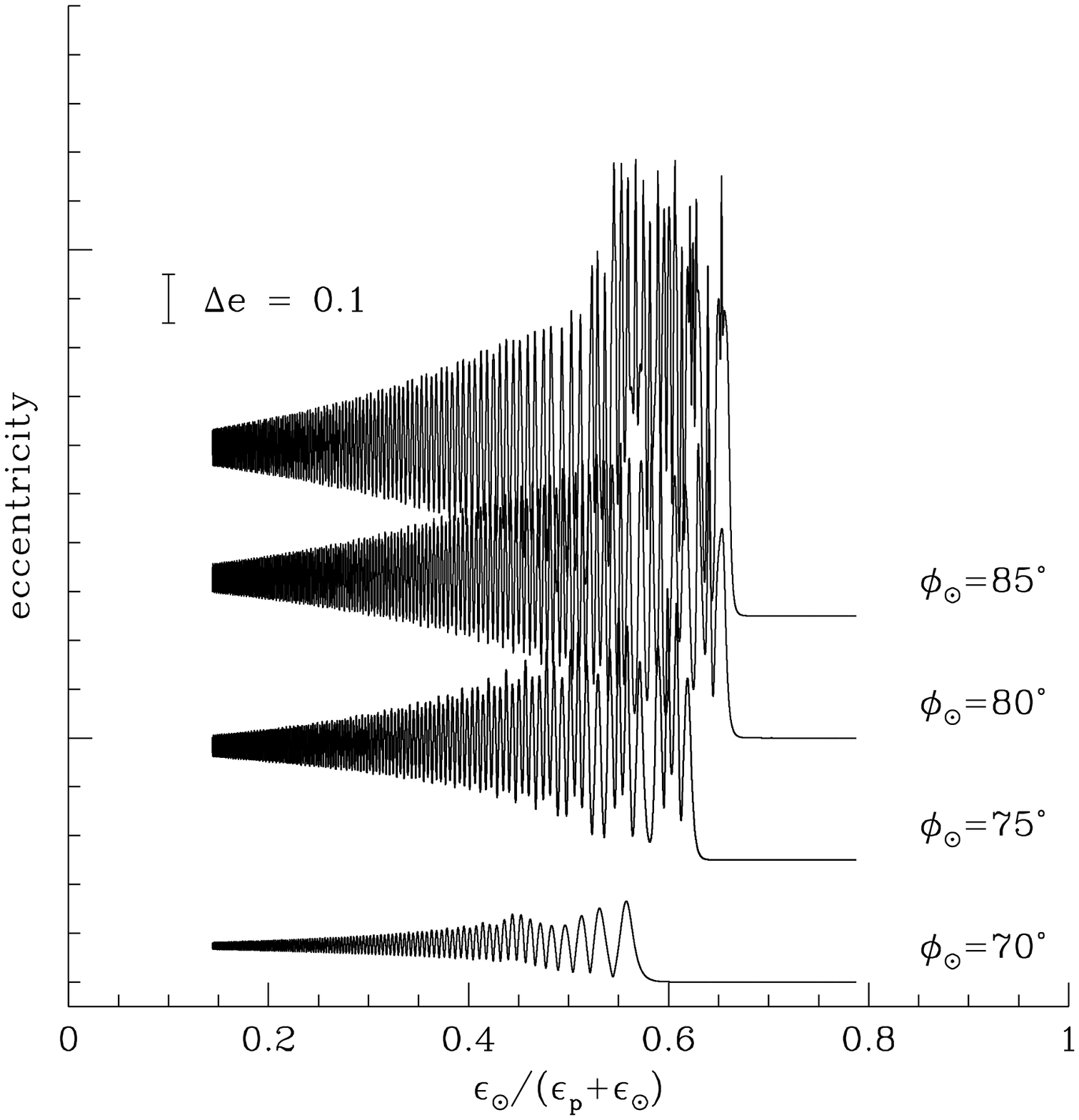}
\caption{Numerical integrations of the equations of motion
  (\ref{eq:milank}). The $y$-coordinate is $\phi_\odot+20e$
  where $e$ is the eccentricity of the solution. Each orbit is begun at
  semimajor axis $a=1.3r_L$ and the semimajor axis then shrinks according to
  $a/r_L=1.3-0.002\tau$ where $\tau$ is defined in equation (\ref{eq:www}). 
  The satellite is initially in the circular Laplace equilibrium except for a
  seed eccentricity of $0.0001$. The $x$-coordinate is
  $\epsilon_\odot/(\epsilon_p+\epsilon_\odot)= a^5/(a^5+r_L^5)$. Four
  integrations are shown, for obliquities
  $\phi_\odot=70^\circ,75^\circ,80^\circ,85^\circ$.} 
\label{fig:orbit}
\end{figure} 

\section{Orbital evolution}

\label{sec:num}

\noindent
Consider the evolution of a satellite on a circular orbit that is slowly
decaying, so the satellite is migrating toward the planet. Let us assume that
the initial semimajor axis is much larger than the Laplace radius $r_L$ (eq.\
\ref{eq:laprad}) and that the orbital plane coincides with the planetary orbit
(i.e., the satellite is in the classical Laplace surface). As outlined at
the end of \S\ref{sec:coco}, if the obliquity $\phi_\odot < 68.875\deg$ we
expect that the satellite will remain in a circular coplanar Laplace equilibrium as it
spirals in. For $68.875\deg < \phi_\odot < 71.072\deg$ we expect that at
semimajor axis $a\simeq 1.2 r_L$ (Fig.\ \ref{fig:unstable}) the satellite will
transfer onto the eccentric coplanar-coplanar equilibrium; its eccentricity will then
grow as it migrates inward, reach a maximum depending on $\phi_\odot$, then shrink
back to zero, at which point ($a\simeq 0.9r_L$) it will rejoin the classical
Laplace surface and spiral into the planet on a circular
orbit. Finally, for $\phi_\odot > 71.072\deg$ the satellite will transfer onto the
eccentric coplanar-coplanar equilibrium, and its eccentricity will grow until its orbit
becomes unstable and begins to execute large
oscillations in eccentricity and inclination

\begin{figure}
\plotone{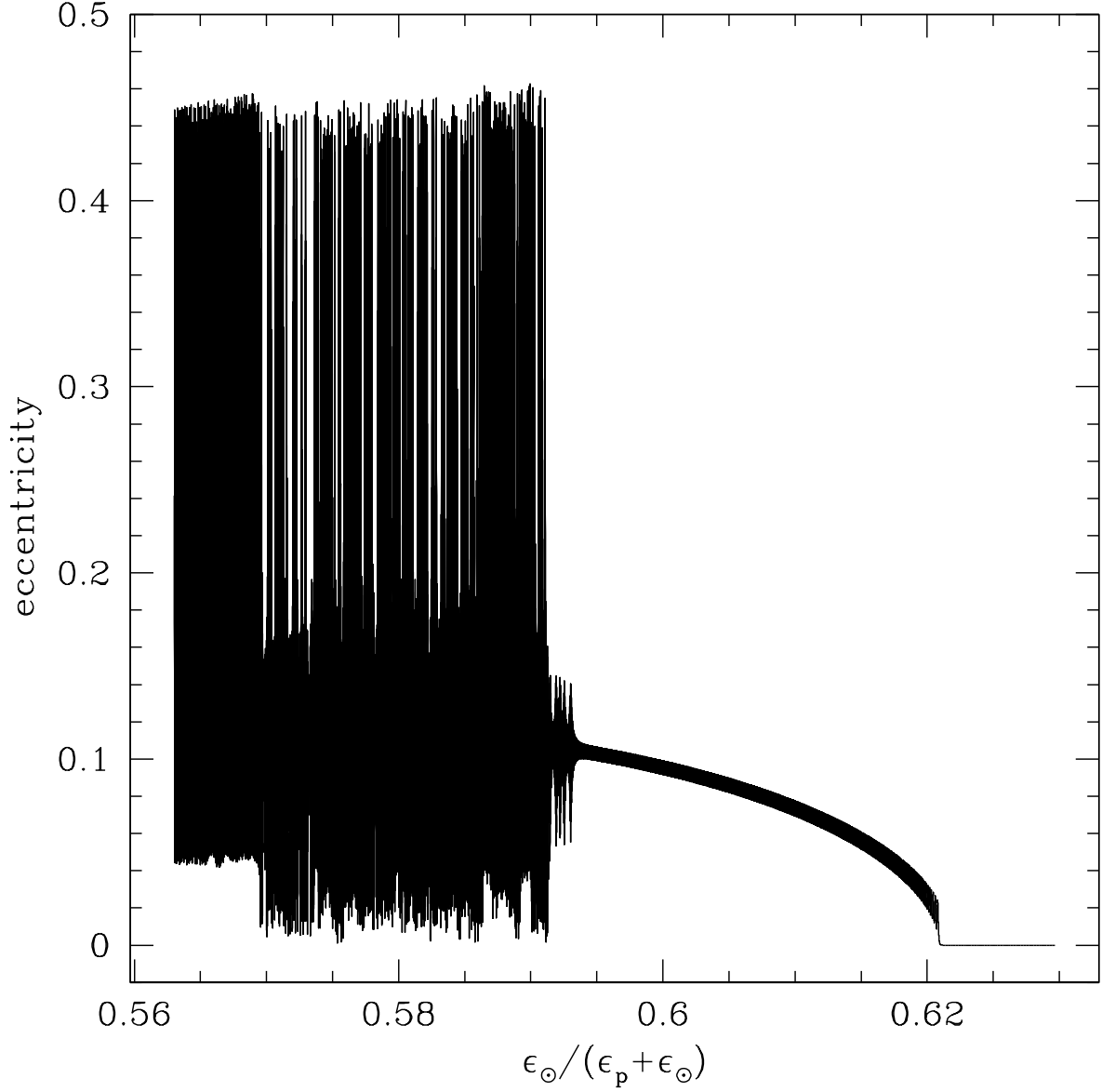}
\caption{Numerical integration of the equations of motion
  (\ref{eq:milank}) as a satellite migrates inward. The obliquity is
  $\phi_\odot=70^\circ$ and the satellite semimajor axis shrinks according to 
  $a/r_L=1.112-0.000002\tau$ where $\tau$ is defined in equation (\ref{eq:www}). 
  The satellite is initially in the circular Laplace equilibrium except for a
  seed eccentricity of $10^{-6}$. The $x$-coordinate is
  $\epsilon_\odot/(\epsilon_p+\epsilon_\odot)= a^5/(a^5+r_L^5)$. The satellite
  initially lies on the classical Laplace surface; at
  $\epsilon_\odot/(\epsilon_p+\epsilon_\odot)=0.621$ it transitions to the
  eccentric coplanar-coplanar sequence; and at
  $\epsilon_\odot/(\epsilon_p+\epsilon_\odot)=0.593$ it transitions to an
  orbit that exhibits large and irregular eccentricity oscillations.}  
\label{fig:orbita}
\end{figure}

Numerical integrations of the equations of motion (\ref{eq:milank})
are shown for a migrating satellite in Figure \ref{fig:orbit}. As
expected, when the initial obliquity $\phi_\odot$ is $75^\circ$,
$80^\circ$, or $85^\circ$ large eccentricity oscillations develop in
the region where the circular coplanar Laplace equilibrium is unstable
(compare Fig.\ \ref{fig:coco}). Smaller oscillations develop in the
case $\phi_\odot=70^\circ$; these are unexpected since the eccentric
coplanar-coplanar sequence is stable at this obliquity, so we would
expect smooth growth and decline in the eccentricity as the semimajor
axis declines, as in Figure \ref{fig:coco}.  Further integrations 
show that the behavior of migrating satellites at this obliquity
depends on the migration time and other parameters; for example, the
orbit shown in Figure \ref{fig:orbita} follows the eccentric
coplanar-coplanar sequence for a while and then rather suddenly jumps
to an orbit with a chaotic appearance that reaches eccentricities as
high as 0.45. These results suggest the presence of large chaotic
regions and perhaps higher order resonances in the phase space, but we
have not yet explored these features.

\section{Applications}

\noindent
Direct applications of these results to the solar system are rather
few. Only Uranus and Pluto have obliquities that exceed the critical
value $\phi_\odot=68.875^\circ$ at which the classical Laplace surface
becomes unstable, and these do not have satellites close to the
unstable range of semimajor axis.

The circular orthogonal Laplace equilibria have been suggested as possible sites for
``polar'' rings around Neptune (see \citealt{dob89} and references therein).
In this case the solar quadrupole potential is much weaker than the potential
from Neptune's satellite Triton. However, the effects of Triton can be modeled
approximately using the formalism we have derived, simply replacing the Sun by
Triton. We find from equation (\ref{eq:laprad}) that the Neptune-Triton
Laplace radius is $r_L=2.15\times 10^{10}\cm=8.54R_p$. Equation (\ref{eq:al})
then implies that circular polar rings would be unstable at
semimajor axes $\gtrsim 7.4R_p$, and beyond this radius polar rings must be
eccentric. This analysis is only approximate, since (i) the Laplace radius
$r_L$ is 61\% of Triton's semimajor axis, $a_T=3.548\times10^{10}\cm$, so the
approximation of Triton's potential by its quadrupole component is poor; (ii)
Triton's orbital axis and Neptune's spin axis both precess around their mutual
invariable plane, whereas we have assumed that they are fixed; this is
probably only a minor correction, since the precession period of $680\yr$ is
much longer than the growth time of the instability,
$(a^3/GM_p)^{1/2}/\lambda=17.3\yr$ at $a=r_L$; (iii) our calculation neglects
collective effects in the ring, such as interparticle collisions, which might
suppress the instability.

Our original motivation for examining this problem was the orbit of
Saturn's satellite Iapetus, which has an eccentricity of 0.028 and an
inclination of $7.5^\circ$ to the Laplace surface. If it formed from a
circumplanetary disk, one might expect Iapetus to have zero
eccentricity and inclination relative to this surface. Thus it appears
that some process has pumped up Iapetus's inclination while leaving
its eccentricity near zero. \cite{ward} pointed out that the shape of
the classical Laplace surface is affected by the mass in the
circumplanetary disk, and suggested that the current orbit of Iapetus
reflects its shape before the disk dispersed. However, this scenario
requires the dispersal of the disk in $\lesssim 10^3$ yr---if the
dispersal were slower, the inclination relative to the Laplace surface
would be an adiabatic invariant and thus would remain near zero.  The
semimajor axis of Iapetus $a=59R_p$ is not far from the Laplace radius
$r_L=48R_p$ (Table 1) so it is natural to wonder whether instabilities
have excited the inclination. However, (i) at Saturn's obliquity of
$26.7^\circ$, all circular orbits in the Laplace surface are stable;
(ii) the instability we have found in the Laplace surface at high
obliquity tends to excite eccentricity, not inclination.

The rich dynamics described in this paper is likely to play a larger role in
scenarios in which the obliquities of the giant planets have changed
substantially since the formation of the solar system \citep[e.g.,][]{tre91}.

A further application is to extrasolar planetary systems. It is
well-known that Kozai oscillations can excite the eccentricity of a
single planet orbiting one member of a binary star system
\citep{koz,hol,for}. Now consider a star that hosts two planets, one a
hot Jupiter, and belongs to a binary system (see \citealt{tak08} for
numerical simulations of such systems). The hot
Jupiter augments the (otherwise negligible) quadrupole moment of the
host star, just as inner satellites augment the quadrupole moment of a
planet (eq.\ \ref{eq:inner}). The resulting Laplace radius
(\ref{eq:laprad}) is \be r_L=2.7\,\hbox{AU }\left(m_J\over 0.001
M_\star\right)^{1/5}\left(a_J\over
0.1\,\hbox{AU}\right)^{2/5}\left(a_\star\over
300\,\hbox{AU}\right)^{3/5}(1-e_\star^2)^{3/10}, \ee where $m_J$ and
$a_J$ are the mass and semimajor axis of the hot Jupiter, and
$M_\star$, $a_\star$, and $e_\star$ are the mass, semi-major axis, and
eccentricity of the binary companion.  If, as we expect, the orbital
axes of the hot Jupiter and the distant companion star are
uncorrelated, 36\% of binary stars will have obliquities that exceed
the critical value $68.875^\circ$ at which instabilities set in at
some semimajor axis; and if planetary migration is common we expect
that many planets now in the habitable zone will have passed through
the unstable region and thereby acquired substantial eccentricities
(cf.\ Fig.\ \ref{fig:orbit}).

\section{Summary}

\noindent
We have examined the orbital dynamics of planetary satellites under the
combined influence of the quadrupole moment from the planet's equatorial bulge
and the tidal field from the Sun.  The Laplace equilibria are orbits in which
the secular evolution due to these forces is zero. They represent
the orbits in which planetary satellites formed from a circumplanetary gas
disk should be found.

Laplace equilibria exist for both circular and eccentric orbits. At a given
semimajor axis, the orbit normals to the circular Laplace equilibria lie along
three orthogonal directions, two of them in the principal plane defined by the
planet's spin axis and the normal to its orbit around the Sun (the ``circular
coplanar'' equilibria). One of the two circular coplanar equilibria is always
unstable. The warped surface swept out by the other circular coplanar
equilibrium orbits as the semimajor axis varies is called the classical
Laplace surface (eq.\ \ref{eq:sola}). The classical Laplace surface coincides
with the equator of the planet at small distances and with the orbital plane
of the planet at large distances. Orbits in the classical Laplace surface are
stable if the planetary obliquity $\phi_\odot<68.875^\circ$, while in the
range $68.875^\circ < \phi_\odot < 90^\circ$ there is a range of semimajor
axes near the Laplace radius $r_L$ (eq.\ \ref{eq:laprad}) in which the surface
is unstable. The third Laplace equilibrium for circular orbits (the
``orthogonal'' or ``polar'' equilibrium) corresponds to orbits that cross over
the planet's pole, and these are stable if and only if the semimajor axis is
less than $2^{-1/5}r_L$ (eq.\ \ref{eq:al}).

At some obliquities and semimajor axes eccentric Laplace equilibria
also exist. In the ``coplanar-coplanar'' equilibria both the
eccentricity and angular-momentum vectors lie in the principal plane
(Figures \ref{fig:stablecc1} and \ref{fig:coco}). The
``orthogonal-coplanar'' equilibria are stable, eccentric polar orbits
that bifurcate from the circular orthogonal equilibria at the
semimajor axis $2^{-1/5}r_L$ where these become unstable. Other
eccentric equilibria exist but are unstable.

The use of the secular equations of motion (\ref{eq:milank}) should be
legitimate so long as the precession times for the eccentricity and
angular-momentum vector are much longer than the orbital period of the
satellite. A sufficient condition for this is that the satellite semimajor
axis is much larger than the planetary radius and much smaller than the
planet's Hill radius (cf.\ Table \ref{tab:outer}). A possible limitation of the secular
equations is that they neglect the evection resonance, where the
apsidal precession rate equals the mean motion of the planet around the Sun.
For the outer planets, the evection resonance typically occurs at $\sim
0.2r_L$. \cite{tw98} have stressed the role of the evection resonance in
the history of the lunar orbit.

Several unanswered questions remain: (i) What is the structure of the
four-dimensional phase space of the equations of motion
(\ref{eq:milank}), and what fraction of the orbits are chaotic? These
issues could be explored through surfaces of section, although the
exploration would be laborious since there is a separate surface of
section for each value of the Hamiltonian, semimajor axis, and
planetary obliquity. (ii) What is the nature of the evolution of
satellites in the classical Laplace surface that migrate into an
unstable region (Fig.\ \ref{fig:orbit})? (iii) What are the shape and
properties of dissipative disks in the range of semimajor axes where
the classical Laplace surface is unstable? Does the dissipation
stabilize the classical Laplace surface? Does the disk follow the
eccentric coplanar-coplanar Laplace surface? Or is there no
steady-state disk? Hints that eccentric
structures can occur in dissipative disks include the eccentric structures
seen in the planetary ring systems of Uranus and Saturn, debris disks around
young stars, and galactic nuclei such as M31. (iv) What is the analog of the
Laplace surface in accretion disks around black holes, where Lense--Thirring
precession rather than a quadrupole potential is the dominant non-Keplerian
perturbation from the central body?

\acknowledgments

We thank Yuri Levin for thoughtful comments. This research was
supported in part by NASA grants NNX08AH83G and NNX08AH24G, and by 
NSF grant AST-0807432.

\appendix

\section{Gauge dependence in the averaged potential}

\label{app:a}

\noindent
The orbit-averaged potential $\Psi(\bfj,\bfe)$ is a function of the six scalar
components of $\bfj$ and $\bfe$ but is only physically meaningful on the 
four-dimensional manifold defined by the
constraints (\ref{eq:const}). Thus $\Psi$ may be replaced by $\Psi+F$, where
the gauge function $F(\bfj,\bfe)$ is arbitrary except that
$F=\hbox{const}$ on the manifold (\ref{eq:const}). We now show that
$F(\bfj,\bfe)$ has no effect on the equations of motion (\ref{eq:mot}), just
as adding a constant to a Newtonian potential $\phi$ has no effect on the
equations of motion $\ddot\bfx=-\bnabla\phi$.  

We work in the coordinates with axes parallel to the unit vectors
$(\bfu,\bfv,\bfn)$ defined just before equation (\ref{eq:avg}). Substituting
$F$ for $\Psi$ in the equations of motion (\ref{eq:mot}) and observing that
$\bfj=j\bfn$, $\bfe=e\bfu$, we have
\begin{eqnarray}
{d\bfj\over d\tau}&=&-j\bfv(\bnabla_\bfj F\cdot\bfu) +j\bfu(\bnabla_\bfj
                     F\cdot\bfv) -e\bfn(\bnabla_\bfe F\cdot\bfv)
                     +e\bfv(\bnabla_\bfe F\cdot\bfn) \nonumber \\ 
{d\bfe\over d\tau}&=&-j\bfv(\bnabla_\bfe F\cdot\bfu)
                     +j\bfu(\bnabla_\bfe F\cdot\bfv) -e\bfn(\bnabla_\bfj
                     F\cdot\bfv) +e\bfv(\bnabla_\bfj F\cdot\bfn).
\label{eq:motf}
\end{eqnarray}
The gauge condition that $F=\hbox{const}$ on the manifold (\ref{eq:const}) is
\be
0=dF=\bnabla_\bfj F\cdot d\bfj +\bnabla_\bfe F\cdot d\bfe\quad\hbox{when}\quad
\bfe\cdot d\bfj+\bfj\cdot d\bfe=0\ \hbox{and}\ \bfj\cdot d\bfj+\bfe\cdot
d\bfe=0.
\label{eq:constr}
\ee
We write $d\bfj$ and $d\bfe$ in terms of the unit vectors  $(\bfu,\bfv,\bfn)$
as
\be
d\bfj= a_u\bfu+a_v\bfv+a_n\bfn,\qquad d\bfe=b_u\bfu+b_v\bfv+b_n\bfn;
\ee
then (\ref{eq:constr}) requires that
\be
0=a_u(\bnabla_\bfj F\cdot\bfu)+
  a_v(\bnabla_\bfj F\cdot\bfv)+
  a_n(\bnabla_\bfj F\cdot\bfn)+
  b_u(\bnabla_\bfe F\cdot\bfu)+
  b_v(\bnabla_\bfe F\cdot\bfv)+
  b_n(\bnabla_\bfe F\cdot\bfn)
\label{eq:aaa1}
\ee
when
\be
ea_u+jb_n=0, \quad ja_n+eb_u=0.
\label{eq:aaa2}
\ee
The six variables $a_u,a_v,a_n,b_u,b_v,b_n$ can be chosen arbitrarily so long
as they satisfy the two constraints (\ref{eq:aaa2}). Thus (\ref{eq:aaa1}) can
only be true if the gauge function $F$ satisfies four conditions:
\be
\bnabla_\bfj F\cdot\bfv=0,\quad 
\bnabla_\bfe F\cdot\bfv=0,\quad 
j\bnabla_\bfj F\cdot\bfu=e\bnabla_\bfe F\cdot\bfn,\quad 
e\bnabla_\bfj F\cdot\bfn=j\bnabla_\bfe F\cdot\bfu.
\ee
With these conditions, equations (\ref{eq:motf}) yield
$d\bfj/d\tau=0$, $d\bfe/d\tau=0$, so the gauge function has no effect on the
equations of motion.

\section{Behavior of circular orbits}

\label{app:b}

\noindent
A circular orbit remains circular under the perturbations in question;
that is, there is a set of solutions of equations (\ref{eq:milank}) in
which $\bfe=0$ for all time. The dynamics of the circular orbit is
governed by the three-dimensional system of equations
\be
\frac{d\bfj}{d\tau} =
\frac{3\epsilon_\odot\,\bfj\cdot\bfn_\odot}{4}\bfj
\times\bfn_\odot + \frac{3\epsilon_p\bfj\cdot
\bfn_p}{2} \bfj\times\bfn_p.
\label{eq:circular}
\ee 
This system has two integrals of motion, $|\bfj|$ (equal to unity on the
manifold corresponding to physical solutions) and the associated
Hamiltonian (eq.\ \ref{eq:ham}, for $e=0$). The system is integrable,
as we now demonstrate by recasting the problem in the form of free rigid-body
dynamics with the appropriate inertia tensor.

First, note that $(\bfj\cdot\bfn_\odot)\,\bfj\times\bfn_\odot=\bfj\times
(\bfj\cdot\bfn_\odot)\bfn_\odot$ can be rewritten in the form $\bfj
\times \bfT_\odot\cdot\bfj$ where
\be
\bfT_\odot = {\left(\begin{array}{lll} n_1^2 & n_1 n_2 & n_1 n_3 \\ n_2
  n_1 & n_2^2 & n_2 n_3 \\ n_3 n_1 & n_3 n_2 & n_3^2 \end{array}
  \right)}_\odot
\ee
is a symmetric tensor built out of the components of $\bfn_\odot=
(n_1,n_2,n_3)_\odot$. The same can be done for the $\bfn_p$ term
leading to an equivalent symmetric tensor $\bfT_p$, built out of the
components of $\bfn_p$. We can thus rewrite equation (\ref{eq:circular}) in
the suggestive form 
\be 
\frac{d\bfj}{d\tau} = \bfj \times \bfT \cdot \bfj\quad\hbox{where}\quad
\bfT = \alpha \bfT_\odot + \beta \bfT_p,
\label{eq:reduced}
\ee
with $\alpha={3\over4}\epsilon_\odot$ and $\beta={3\over2}\epsilon_p$.
Following this rearrangement, the program is straightforward: we find
the matrix that diagonalizes the symmetric tensor $\bfT$, identifying
principal directions (actually the directions of the circular Laplace
equilibria discussed in \S\ref{sec:cop}), and the principal
values, which will decide the stability of equilibria and the global
phase-space topology around them. To fix things, it is simplest to
take $\bfn_p=(1,0,0)$, and
$\bfn_p=(\cos\phi_\odot,\sin\phi_\odot,0)$ (obliquity $\phi_\odot$),
leading to: 
\be
\bfT = \left(\begin{array}{lll} 
  \alpha\cos^2\phi_\odot + \beta & \alpha\cos\phi_\odot\sin\phi_\odot
& 0 \\
  \alpha\cos\phi_\odot\sin\phi_\odot & \alpha\sin^2\phi_\odot & 0 \\
  0 & 0 & 0 \end{array} \right).
\ee
The eigenvalues of $\bfT$ are
\begin{eqnarray}
t_1 & = & 0 \nonumber \\
t_2 & = &\frac{\alpha+\beta-\sqrt{(\alpha + \beta)^2 - 4\alpha
  \beta\,\sin^{2}\phi_\odot}}{2} \nonumber \\
t_3 & = & \frac{\alpha + \beta + \sqrt{(\alpha + \beta)^2 - 4\alpha
  \beta\,\sin^{2}\phi_\odot}}{2},
\end{eqnarray}
with $t_1=0 < t_2 < t_3$ ($t_2 = 0$ for $\phi_\odot=0$, an
ignorable case). We can then solve for the eigenvectors $\bft^{(1)}$,
$\bft^{(2)}$, $\bft^{(3)}$ of $\bfT$, the principal axis
directions. Since $\bfT$ is symmetric its eigenvectors are
orthogonal. Moreover the eigenvectors can be chosen to be orthonormal
and to form a right-handed triad
($\bft^{(1)}=\bft^{(2)}\times\bft^{(3)}$). In particular,
$\bft^{(1)}=(0,0,\pm1)$ (assuming, as usual, that $\epsilon_p$,
$\epsilon_\odot$, and $\sin\phi_\odot$ are non-zero), that is,
$\bft^{(1)}$ is perpendicular to the principal plane containing $\bfn_p$
and $\bfn_\odot$, while $\bft^{(2)}$ and $\bft^{(3)}$ lie in the
principal plane.

Let $\bfR$ be the orthogonal transformation associated with these
vectors, the transformation that diagonalizes $\bfT$. The columns of
$\bfR$ are the orthonormal eigenvectors of $\bfT$, so
$R_{ij}=t^{(j)}_i$. We now rotate to the principal-axis coordinate
system, by setting $\bfj = \bfR\cdot\bfJ$. Equation
(\ref{eq:reduced}) turns into 
\begin{eqnarray}
\frac{d J_1}{dt} & = &  (t_3 - t_2) J_2 J_3 \nonumber \\
\frac{d J_2}{dt} & = &  (t_1 - t_3) J_1 J_3 = - t_3 J_1 J_3 \nonumber \\
\frac{d J_3}{dt} & = &  (t_2 - t_1) J_1 J_2 = t_2 J_1 J_2,
\label{eq:eomjt}
\end{eqnarray}
exactly what one would obtain for the asymmetrical top in Euler's formulation,
with $1/t_i$ standing for $I_i$, the principal moments of inertia
\citep{ll}. As in the case of the rigid body, 
vectors that lie along the principal axes (parallel and
anti-parallel) are stationary states of the dynamics: $\bfJ=(\pm 1,0,0)$,
$(0,\pm 1,0)$, and $(0,0,\pm 1)$.

The equilibria along $\bft^{(1)}$ are linearly stable, with linear
oscillation frequency $\Omega_1=\sqrt{t_2 \, t_3}$ ($t_2$ and $t_3$
are both positive); equilibria along $\bft^{(3)}$ are also stable,
with linear oscillation frequency $\Omega_2=\sqrt{(t_3 - t_2)\,t_3}$
($t_3 > t_2 > 0$); equilibria along $\bft^{(2)}$ are linearly unstable
with growth rate $\sqrt{t_2(t_3-t_2)}$. Thus $\bft^{(3)}$ and
$\bft^{(2)}$ describe the directions of the angular-momentum vectors
for the linearly stable and unstable circular coplanar Laplace equilibria,
while $\bft^{(1)}$ describes the circular orthogonal Laplace equilibrium, which
is stable (recall that we consider only circular orbits in this
Appendix, so the stability properties described here do not include
the eccentricity instabilities discussed in \S \ref{sec:stabc}).

We now examine the nonlinear dynamics. The Hamiltonian is the
restriction of $\Psi_p+\Psi_\odot$ (eq.\ \ref{eq:psips}) to circular
orbits, $e=0$. In the present notation this can be written
\be
H_c=-\half\bfj^{\rm T}\cdot\bfT\cdot\bfj=-\half\sum_{i=1}^3
t_iJ_i^2=-\half(t_2J_2^2+t_3J_3^2),
\ee
where the superscript ``T'' denotes transpose. If it is not clear
already, note that $H_c$
generates Lie-Poisson dynamics of $\bfJ$ according to:
\be
\frac{d\bfJ}{d\tau} = - \bfJ \times \bnabla_\bfJ H_c.
\ee 
There are two integrals of motion, $H_c$ itself and the magnitude
$J^2=J_1^2+J_2^2+J_3^2=1$. Thus trajectories lie on the intersection
of the angular momentum sphere and the elliptical cylinder
representing the energy surface (as opposed to the ellipsoid with axes
$1/I_i$ that is involved in the solution of the asymmetrical top). The
largest allowable energy is $H_{c,\rm max}=0$ and the smallest is
$H_{c,\rm min} = -\half t_3$. Starting at the minimum energy,
the cylinder intersects the sphere at two points, corresponding to
equilibria along $\bft^{(3)}$.  Increasing the energy, the cylinder
intersects the sphere in closed curves around these equilibria. This
remains so, till we reach the critical energy $H_c=-\half t_2$, at which
point the cylinder is tangent to the sphere at $\pm\bft^{(2)}$ as it
intersects it along the limiting curves, separatrices, 
that separate the stable librations around $\bft^{(3)}$ from the
stable librations around $\bft^{(1)}$. Increasing further, the
cylinder now intersects the sphere at two librating curves around $\pm
\bft^{(1)}$, until it reduces to a needle piercing the sphere at $\pm
\bft^{(1)}$ when $H_c = 0$. 

Orbit shapes are easy to obtain in projection. For trajectories
librating around the $\bft^{(3)}$ equilibrium, take the ratio of the
$J_1$ and $J_2$ equations and integrate (or eliminate $J_3^2$ between
the energy and angular-momentum integrals) to get
\be
(t_3 - t_2)J_2^2+t_3 J_1^2 = \hbox{const}=t_3+2H_c,\qquad 
-t_3 \leq 2 H_c < -t_2.
\ee
At the limiting value, $H_c = -\half t_2$, this defines the separatrices in
projection. Similarly we can obtain trajectories librating around the 
$\bft^{(1)}$ equilibrium. Of course the equations of motion
(\ref{eq:eomjt}) can be solved explicitly, following the methods used
for the asymmetrical top \citep{ll}.

Incidentally, if the system is subject to a process that dissipates
energy while conserving angular momentum, it will decay towards the
minimum energy state, which lies along $\pm\bft^{(3)}$, the direction
normal to the circular coplanar Laplace equilibrium. This is often interpreted
to mean that the classical Laplace surface, at a minimum of $H_c$, is
secularly stable while the circular orthogonal Laplace equilibrium at a maximum
of $H_c$ is secularly unstable. However, most dissipative processes
also affect the eccentricity and semimajor axes of the orbits, so the
evolution of the orientation of the angular-momentum vector due to
dissipation cannot be treated in isolation. In particular,
\citet{dob89} argue in the context of possible polar rings around
Neptune that the circular orthogonal Laplace equilibria are secularly stable. 

\section{Classification of eccentric Laplace equilibria} 

\label{app:c}

\noindent
We investigate the properties of Laplace equilibria (stationary solutions of
eqs.\ \ref{eq:milank}) with non-zero eccentricity. As usual we assume that
$\epsilon_p>0$ and that $\bfn_\odot$ is neither parallel,
antiparallel, nor perpendicular to $\bfn_p$. Recall that $|\bfj|=(1-e^2)^{1/2}$.

Take the scalar product of the first of equations (\ref{eq:milank}) with
$\bfn_\odot$. The first two terms on the right side vanish, and since
$d\bfj/d\tau=0$ in a stationary solution and $\epsilon_p\not=0$ we must have
either 
\be
\hbox{(a)}\ \bfn_\odot\cdot(\bfj\times\bfn_p)=0,\qquad\hbox{or}\qquad
\hbox{(b)}\ \bfj\cdot\bfn_p=0.
\ee

First consider case (a). Take the scalar product of the second of equations
(\ref{eq:milank}) with $\bfe$. The first, third, and fourth terms on the right
side vanish, and $\epsilon_\odot\not=0$, so we must have either 
\begin{eqnarray}
\hbox{(aa)}\ \bfn_\odot\cdot(\bfj\times\bfn_p)=0\quad&\hbox{and}&\quad
     \bfe\cdot(\bfj\times\bfn_\odot)=0, \qquad\hbox{or}\nonumber \\
\hbox{(ab)}\ \bfn_\odot\cdot(\bfj\times\bfn_p)=0\quad&\hbox{and}&\quad
     \bfe\cdot\bfn_\odot=0. 
\end{eqnarray}
Case (aa) requires that $\bfj$, $\bfe$, $\bfn_\odot$, and $\bfn_p$ are all
coplanar, that is, that $\bfj$ and $\bfe$ lie in the principal plane defined
by $\bfn_\odot$ and $\bfn_p$. We call this the ``coplanar-coplanar'' solution
since $\bfj$ and $\bfe$ lie in the
principal plane. Case (ab) requires that $\bfj$, $\bfn_\odot$, and $\bfn_p$
are coplanar and that $\bfe$ is perpendicular to $\bfn_\odot$; moreover by
definition $\bfe$ is perpendicular to $\bfj$. Hence $\bfj$ lies in the
principal plane and $\bfe$ is normal to this plane; we call this the
``coplanar-orthogonal'' solution. 

Now consider case (b). Take the scalar product of the first of equations
(\ref{eq:milank}) with $\bfe$. The second and third terms on the right side
vanish, so either
\begin{eqnarray}
\hbox{(ba)}\ \bfj\cdot\bfn_p=0\quad&\hbox{and}&\quad \bfj\cdot\bfn_\odot=0, 
\quad\hbox{or}\nonumber \\
\hbox{(bb)}\ \bfj\cdot\bfn_p=0\quad&\hbox{and}&\quad
     \bfe\cdot(\bfj\times\bfn_\odot)=0.
\end{eqnarray}
In case (ba), the first of equations (\ref{eq:milank}) implies that
$(\bfe\cdot\bfn_\odot)\bfe\times\bfn_\odot=0$. Hence either
\begin{eqnarray}
\hbox{(baa)}\ \bfj\cdot\bfn_p=0\quad&\hbox{and}&\quad
\bfj\cdot\bfn_\odot=0\quad\ \hbox{and}\quad\
\bfe\cdot\bfn_\odot=0,\quad\hbox{or}\nonumber\\ 
\hbox{(bab)}\ \bfj\cdot\bfn_p=0\quad&\hbox{and}&\quad
\bfj\cdot\bfn_\odot=0\quad\ \hbox{and}\quad\ \bfe\times\bfn_\odot=0.
\end{eqnarray}
In case (baa), $\bfj$ is perpendicular to the principal plane, and $\bfe$ lies
in the principal plane at right angles to $\bfn_\odot$; we call this the
``orthogonal-coplanar'' solution. In case (bab) $\bfe$ is parallel or
antiparallel to $\bfn_\odot$ so we can write $\bfe=\pm e\bfn_\odot$ and the
second of equations (\ref{eq:milank}) reduces to 
\be
  3\epsilon_\odot+{\epsilon_p\over(1-e^2)^{5/2}}=0,
\ee
which has no solution since $\epsilon_\odot,\epsilon_p>0$.

In case (bb), $\bfe$, $\bfj$, and $\bfn_\odot$ lie in the same plane, and
the first of equations (\ref{eq:milank}) reduces to
\be
(\bfj\cdot\bfn_\odot)\bfj\times\bfn_\odot =
5(\bfe\cdot\bfn_\odot)\bfe\times\bfn_\odot. 
\label{eq:ggghhh}
\ee
Let $\bfe$ and $\bfj$ define the positive $x$- and $y$-axes of a Cartesian
coordinate system and in this system let
$\bfn_\odot=(\cos\psi,\sin\psi)$. Then equation (\ref{eq:ggghhh}) becomes
$(1+4e^2)\sin\psi\cos\psi=0$. Hence either $\sin\psi=0$ or $\cos\psi=0$. If
$\sin\psi=0$, $\bfe$ is parallel or antiparallel to $\bfn_\odot$, and we
return to case (bab), which has no solution. If $\cos\psi=0$, then $\bfj$ is
parallel or antiparallel to $\bfn_\odot$, so $\bfn_\odot\cdot\bfn_p=0$, a
special case (planetary obliquity of $90^\circ$) that we have already excluded
from consideration. 

Thus the only eccentric Laplace equilibria are case (aa), in which
both $\bfj$ and $\bfe$ lie in the principal plane formed by $\bfn_p$ and
$\bfn_\odot$ (coplanar-coplanar equilibrium); case (ab), in which $\bfj$ lies
in the principal plane and $\bfe$ is normal to this plane (coplanar-orthogonal
equilibrium); or case (baa), in which $\bfe$ lies in the principal plane and
$\bfj$ is normal to the plane (orthogonal-coplanar equilibrium).

\end{document}